\documentclass[12pt,aps]{revtex4}%
\usepackage{amssymb}
\usepackage{amsmath}
\usepackage{amsfonts}
\usepackage{graphicx}%
\begin{document}
\title{Not So Classical Mechanics -- Unexpected Symmetries of Classical Motion}
\author{James T. Wheeler}
\affiliation{Department of Physics, Utah State University, Logan, UT 84322-4415}

\begin{abstract}
A survey of topics of recent interest in Hamiltonian and Lagrangian dynamical
systems, including accessible discussions of regularization of the central
force problem; inequivalent Lagrangians and Hamiltonians; constants of central
force motion; a general discussion of higher-order Lagrangians and
Hamiltonians with examples from Bohmian quantum mechanics, the Korteweg-de
Vries equation and the logistic equation; gauge theories of Newtonian
mechanics; classical spin, Grassmann numbers, and pseudomechanics.

\end{abstract}
\maketitle

\section{Introduction}

The study of classical mechanics is vast and ancient. Therefore, this
collection of results, observations and questions necessarily omits most of
the field and probably misses a number of older references even for topics
covered in detail. We focus principally on issues of symmetry and subjects
(old and new) which have appeared in the literature within recent decades. Nor
should it be thought that we provide a complete survey of even the results we
do discuss. Instead, our references for each topic are probably sufficient for
the interested reader to gain a foothold on the relevant research.

Of course, the list of topics we do not examine is extensive. Certain topics
such as nonlinear dynamics receive only a brief mention as an example of a
higher order system in Section 5. We have chosen to omit any discussion of
electromagnetism while touching on special and general relativity only to
provide examples. Since our presentation is intended for a broad audience, we
have also avoided the large body of formal work. Thus, while the formal study
of symplectic manifolds, Kahler manifolds, Poincar\'{e} sections and so on
make heavy use of modern differential geometry and field theory techniques,
little mention is made of progress in these directions.

What remains is nonetheless filled with fascinating and diverse surprises in a
field often mistaken to be complete. Thus, what we do cover is a wide array of
topics ranging from the Kepler problem to supersymmetry. The unifying theme,
if there is one, is the occurrence of unexpected and surprising symmetries in
classical physics, and especially in Lagrangian and Hamiltonian dynamics.
Though we treat a few topics simply because there is recent reference to them
in the literature, most of the topics concern symmetry in one way or another.
The uses vary greatly, from the use of anticommuting numbers to the rotation
group, from unusual constants of motion of the Kepler problem to the infinite
hierarchy of constants of motion of the Korteweg-de Vries (KdV) equation. An
additional guiding principle has been to treat topics that may not be familiar
to many readers.

Curiously, quantum mechanics and quantum field theory have had a strong impact
on current work in classical physics. As a result, a brief discussion of
quantum mechanics appears in our examination of inequivalent Lagrangians in
Section 3 and Bohmian quantum mechanics is discussed in Section 5. Further
connections between quantum and classical mechanics are suggested in our
treatment of Lagrangian and Hamiltonian dynamics as gauge theories in Section
6. Finally, Section 7 owes its entire existence to insights from
supersymmetric quantum field theory.

The organization of the paper is simultaneously from old to new and from easy
to difficult. Section 2 provides a warm-up exercise with some new thoughts on
an old topic -- the regularization of the central force problem. From there we
move gradually to more recent and more mathematically challenging questions.
In Section 3 we discuss inequivalent Lagrangians and Hamiltonians, a topic
which begins with Lie and Dirac (if not earlier) and which received
considerable new input in the 70s. Through the same period, old knowledge
resurfaced with the rediscovery of the Laplace-Runge-Lenz and Hamilton
vectors. A derivation of these rediscovered constants of the motion is given
in Section 4, using a technique based on an old theorem. While the theorem
will no doubt be familiar to mathematicians, its simple application to finding
constants of the motion does not appear in classical physics textbooks.

Moving to more active areas of current interest, we look at the occurrence of
higher order differential equations in classical physics. After a brief
general introduction to these systems at the beginning of Section 5 is an
example of such equations -- Bohmian quantum mechanics.

The final two Sections deal with truly contemporary insights. In Section 6, a
development of both Lagrangian and Hamiltonian dynamics as gauge theories
shows an interesting new connection between classical physics and conformal
symmetry. Then, in Section 7, two further developments of field theory --
spinors and anticommuting variables -- are discussed in the context of
particle mechanics. The KdV equation and the approach to chaos are treated in Appendices.

Before embarking, a few general comments are in order. First, observe that
each section below is essentially independent of the others. Each section has
its own brief introduction and references. Note that the end of most sections
we have tried to provide a few stimulating questions. These questions do not
reflect any consensus thinking and should not be taken to be the definitive
puzzles facing the field. Rather, they are suggestions of some directions
which might or might not prove fruitful. Finally, it should be noted that
where derivations are given without citation, we have produced original
calculations. We make no further mention of this fact since it is probable
that many or all of these calculations have already appeared somewhere within
the last few hundred years!

\section{Regularization of the central force problem}

We begin with some of the oldest problems of classical physics. In this
Section and the next, we explore some interesting features of the Kepler
problem and other central force motion. In this Section, we examine
regularizations of the Kepler problem. In the next section we present a
technique for finding constants of the motion \cite{TjiangSutanto}, then, as
an example, use the technique to find some recently rediscovered constants of
the Kepler problem \cite{Munoz}.

Regularizations of dynamical problems are transformations that turn the
equations of motion into a simpler or less singular mechanical problem. Euler
\cite{Euler} and Levi-Civita \cite{Levi-Civitö} produced one- and
two-dimensional regularizations, respectively, of the Kepler problem. These
regularizations turn the Kepler/Coulomb equation of motion into an isotropic
oscillator. It is not surprising that this is possible, because the
transformations are time-dependent. Indeed, using similar transformations, it
is possible locally to turn any central force problem into the isotropic
oscillator. We present a general proof of this claim below. The Euler and
Levi-Civita results are special cases.

There are some recent discussions in the literature extending these
regularizations. Since the Levi-Civita result makes use of a complex variable,
some authors have explored the idea that the use of a vector space which is
also a number field gives insights into the problem. Thus, Kustaanheimo and
Stiefel (\cite{Kustaan1964},\cite{KustaanStiefel1965},\cite{Stiefel}) give a
quaternionic transformation from the $3$-dim Kepler problem to a constrained
$4$-dim isotropic oscillator, thereby showing that bounded Kepler orbits have
an underlying $SO(4)$ symmetry. Bartsch \cite{Bartsch} writes the
Kustaanheimo-Steifel result in terms of Hestenes' geometric algebra
\cite{Hestenes}. Such use of quaternionic, Clifford or Grassmann variables
(see below) often extends, or streamlines, the presentation of classical results.

However, it seems unlikely that number fields are necessary to transform the
Kepler problem into the oscillator. If that were the case, we would expect
regularization to be possible only using real, complex, quaternionic or
octonionic variables and therefore only to occur in dimensions less than or
equal to eight. But since both the Kepler problem and the isotropic oscillator
are inherently two dimensional, the Levi-Civita solution should suffice in any
higher dimension as well. Our generalized solution below demonstrates this to
be the case.

\subsection{Higher dimensions}

Consider the general central force motion in any dimension $d\geq2.$ We begin
from the action
\[
S=\int dt\left[  \frac{1}{2}m\frac{dx_{i}}{dt}\frac{dx_{i}}{dt}-V\left(
r\right)  \right]
\]
where $r=\sqrt{x_{i}x_{i}}.$ It follows that
\[
m\frac{d^{2}x_{i}}{dt^{2}}=-V^{\prime}\frac{x_{i}}{r}%
\]
We first compute the total angular momentum
\begin{align*}
M_{ij}  &  =x_{i}p_{j}-x_{j}p_{i}\\
&  =m\left(  x_{i}\dot{x}_{j}-x_{j}\dot{x}_{i}\right)
\end{align*}
This is conserved, since
\begin{align*}
\frac{d}{dt}M_{ij}  &  =m\frac{d}{dt}\left(  x_{i}\dot{x}_{j}-x_{j}\dot{x}%
_{i}\right) \\
&  =m\left(  x_{j}\frac{d^{2}x_{k}}{dt^{2}}-x_{k}\frac{d^{2}x_{j}}{dt^{2}%
}\right) \\
&  =-\frac{V^{\prime}}{r}\left(  x_{j}x_{k}-x_{k}x_{j}\right) \\
&  =0.
\end{align*}

To prove from this that the motion lies in a plane, let $\mathbf{x}_{0}$ and
$\mathbf{v}_{0}$ be the initial position and velocity. Then the angular
momentum is
\[
M_{ij}=x_{0i}v_{0j}-x_{0j}v_{0i}\neq0
\]
Let $\mathbf{w}^{\left(  a\right)  },$ $a=1,\ldots,n-2,$ be a collection of
vectors perpendicular to the initial plane
\begin{align*}
P  &  =\left\{  \mathbf{v}=\alpha\mathbf{x}_{0}+\beta\mathbf{v}_{0}\left\vert
\forall\alpha,\beta\right.  \right\} \\
\mathbf{w}^{\left(  a\right)  }\mathbf{v}  &  =0
\end{align*}
so that the set $\left\{  \mathbf{x}_{0},\mathbf{v}_{0},\mathbf{w}^{\left(
a\right)  }\right\}  $ forms a basis. Then, for all $a$%
\[
w_{i}^{\left(  a\right)  }M_{ij}=0.
\]
Now, at any time $t,$ $M_{ij}$ is given by
\[
M_{ij}=m\left(  x_{i}v_{j}-x_{j}v_{i}\right)
\]
and since $M_{ij}$ is constant we still have
\begin{align*}
0  &  =w_{i}^{\left(  a\right)  }m\left(  x_{i}v_{j}-x_{j}v_{i}\right) \\
0  &  =\left(  \mathbf{w}^{\left(  a\right)  }\cdot\mathbf{x}\right)
\mathbf{v}-\mathbf{x}\left(  \mathbf{w}^{a}\cdot\mathbf{v}\right)
\end{align*}
Suppose, for some $a_{0},$ that
\[
\mathbf{w}^{\left(  a_{0}\right)  }\cdot\mathbf{x}\neq0
\]
Then
\[
\mathbf{v}=\mathbf{x}\left(  \frac{\mathbf{w}^{a_{0}}\cdot\mathbf{v}%
}{\mathbf{w}^{a_{0}}\cdot\mathbf{x}}\right)
\]
and $M_{ij}$ is identically zero, in contradiction to its constancy.
Therefore, we conclude
\[
\mathbf{w}^{\left(  a\right)  }\cdot\mathbf{x}=0
\]
for all $a.$ A parallel argument shows that
\[
\mathbf{w}^{\left(  a\right)  }\cdot\mathbf{v}=0
\]
for all $a,$ so the motion continues to lie in the original plane.

Now we choose polar coordinates in the plane of motion, and the problem
reduces to two dimensions. We next need to deal with the presence of angular
momentum. With coordinates $x^{\left(  a\right)  }$ in the $\mathbf{w}%
^{\left(  a\right)  }$ directions, the central force equations of motion are
\begin{align*}
m\frac{d^{2}x^{\left(  a\right)  }}{dt^{2}}  &  =0\\
m\left(  \frac{d^{2}r}{dt^{2}}-r\frac{d\varphi}{dt}\frac{d\varphi}{dt}\right)
&  =-V^{\prime}\left(  r\right) \\
\frac{d\left(  mr^{2}\dot{\varphi}\right)  }{dt}  &  =0
\end{align*}
We choose $x^{\left(  a\right)  }=0,$ and set $L=mr^{2}\dot{\varphi}=$
constant. Eliminating $\dot{\varphi},$ these reduce to the single equation
\begin{equation}
m\frac{d^{2}r}{dt^{2}}-\frac{M^{2}}{mr^{3}}=-V^{\prime}\left(  r\right)
\label{Radial equation}%
\end{equation}

Notice that now any transform of $r$ will change the required form of the
angular momentum term. What works to avoid this is to recombine the angular
momentum and force terms. We again start with
\[
r=f(u),\text{ \ }\frac{d}{dt}=\frac{1}{f^{\prime}}\frac{d}{d\tau}.
\]
Then eq.(\ref{Radial equation}) becomes
\[
\frac{1}{f^{\prime}}\frac{d}{d\tau}\left(  \frac{1}{f^{\prime}}f^{\prime}%
\frac{du}{d\tau}\right)  -\frac{M^{2}}{m^{2}f^{3}}=-\frac{dV}{df}[f\left(
u\right)  ].
\]
Rearranging, we have
\begin{align*}
\frac{d^{2}u}{d\tau^{2}} &  =\frac{M^{2}f^{\prime}}{m^{2}f^{3}}-f^{\prime
}\frac{dV}{df}[f\left(  u\right)  ]\\
&  =\frac{M^{2}f^{\prime}}{m^{2}f^{3}}-\frac{df}{du}\frac{dV}{df}[f\left(
u\right)  ]\\
&  =\frac{M^{2}}{m^{2}f^{3}}\frac{df}{du}-\frac{dV}{du}[f\left(  u\right)  ]
\end{align*}
To obtain the isotropic harmonic oscillator we require the combined angular
momentum and force terms to give the needed expression:
\[
\frac{M^{2}}{m^{2}f^{3}}\frac{df}{du}-\frac{dV}{du}[f\left(  u\right)
]=\frac{\tilde{M}^{2}}{m^{2}u^{3}}-ku
\]
Integrating,
\begin{equation}
\frac{M^{2}}{2m^{2}f^{2}}+V[f\left(  u\right)  ]=\frac{\tilde{M}^{2}}%
{2m^{2}u^{2}}+\frac{1}{2}ku^{2}+\frac{c}{2}.\label{Integrated eq of motion}%
\end{equation}
If we define
\[
g\left(  f\right)  \equiv\frac{M^{2}}{2m^{2}f^{2}}+V[f\left(  u\right)  ]
\]
the required function $f$ is
\[
f=g^{-1}\left(  \frac{\tilde{M}^{2}}{2m^{2}u^{2}}+\frac{1}{2}ku^{2}+\frac
{c}{2}\right)  .
\]
Substituting this solution into the equation of motion, we obtain the equation
for the isotropic oscillator,
\[
m\frac{d^{2}u}{dt^{2}}-\frac{\tilde{M}^{2}}{mu^{3}}=-ku.
\]
Therefore, every central force problem is locally equivalent to the isotropic
harmonic oscillator. Of course, the same result follows from Hamilton-Jacobi
theory, since \textit{every} pair of classical systems with the same number of
degrees of freedom are related by some time-dependent canonical transformation.

The solution takes a particularly simple form for the Kepler problem,
$V=-\alpha/r$. In this case, eq.(\ref{Integrated eq of motion}) becomes
\[
\frac{M^{2}}{2m^{2}f^{2}}-\frac{\alpha}{f}-\left(  \frac{\tilde{M}^{2}}%
{2m^{2}u^{2}}+\frac{1}{2}ku^{2}+\frac{c}{2}\right)  =0
\]
Solving the quadratic for $1/f,$ we take the positive solution
\begin{align*}
\frac{1}{f}  &  =\frac{m^{2}}{M^{2}}\left[  \alpha+\sqrt{\alpha^{2}%
+\frac{M^{2}}{m^{2}}\left(  \frac{\tilde{M}^{2}}{m^{2}u^{2}}+ku^{2}+c\right)
}\right] \\
&  =\frac{\alpha m^{2}}{M^{2}}\left[  1+\frac{M}{\alpha mu}\sqrt
{ku^{4}+\left(  c+\frac{\alpha^{2}m^{2}}{M^{2}}\right)  u^{2}+\frac{\tilde
{M}^{2}}{m^{2}}}\right]  .
\end{align*}
There is also a negative solution.

We may choose $c$ to complete the square under the radical and thereby
simplify the solution. Setting
\[
c=\frac{2\sqrt{k}M\tilde{M}}{m}-\frac{\alpha^{2}m^{2}}{M^{2}}%
\]
the positive solution for $f$ reduces to
\[
\frac{1}{f}=\frac{\alpha m^{2}}{M^{2}}+m\sqrt{k}u+(\tilde{M}/M)\frac{1}{u}%
\]
or
\[
f=\frac{u}{(m\sqrt{k})u^{2}+(\alpha m^{2}/M^{2})u+\tilde{M}/M}.
\]
The zeros of the denominator never occur for positive $u,$ so the
transformation $f$ is regular in the Kepler case. The regularity of the Kepler
case is not typical -- it is easy to see that the solution for $f$ may have
many branches. The singular points of the transformation in these cases should
give information about the numbers of extrema of the orbits, the stability of
orbits, and other global properties. The present calculation may provide a
useful tool for studying these global properties in detail. The problem of
global properties of orbits remains open -- power law forces have been
examined \cite{RayShamanna}, but more complicated potentials allow arbitrarily
many extrema. For example, the potential
\[
V=\alpha\left(  r-r_{0}\right)  ^{2p}%
\]
gives an effective potential
\[
V_{eff}=\frac{M^{2}}{2mr^{2}}+\alpha\left(  r-r_{0}\right)  ^{2p}%
\]
Straightforward perturbation about circular orbits shows that, for arbitrary
fixed angular momentum $M,$ the frequency of radial oscillations may be
increased without bound by increasing $p.$ Such closed orbits will have
arbitrarily many extrema.

\subsection{Euler's regularization}

Essential features of the regularizing transformation are evident even in the
$1$-dim case. The Euler solution uses the substitutions
\[
x=-u^{-2},\text{ \ \ }\frac{d}{dt}=u^{3}\frac{d}{d\tau}%
\]
to turn the $1$-dim Kepler equation of motion into the $1$-dim harmonic
oscillator. Before moving to a proof for the general $n$-dim case, we note
that more general transformations are possible in the $1$-dim case. Beginning
with the equation of motion,
\[
m\frac{d^{2}x}{dt^{2}}=-\frac{\alpha}{x^{2}}%
\]
let
\[
x=f\left(  u\right)  ,\text{ \ \ }\frac{d}{dt}=\frac{1}{f^{\prime}}\frac
{d}{d\tau}.
\]
Then
\[
\dot{x}=f^{\prime}\frac{du}{dt}=\frac{du}{d\tau}%
\]
so the equation of motion becomes
\[
m\frac{d^{2}u}{d\tau^{2}}=-\frac{\alpha f^{\prime}}{f^{2}}.
\]
Now let $V\left(  u\right)  $ be any potential. Demanding
\[
V^{\prime}=\alpha\frac{f^{\prime}}{f^{2}}%
\]
we integrate to find
\[
f=-\frac{\alpha}{V\left(  u\right)  -V_{0}}.
\]
With this choice for $f,$ the equation of motion becomes simply
\[
m\frac{d^{2}u}{d\tau^{2}}=-V^{\prime}.
\]
Notice that $u$ is not necessarily a monotonic function of $x$ so the
transformation at zeros of $V^{\prime}$ may be singular. We will not deal with
such global issues here.

In higher dimensions the regularizing transformation is complicated by the
presence of angular momentum. Still, the general proof is similar, involving a
change of both the radial coordinate and the time. Once again, more general
potentials can be treated. To begin, we eliminate the angular momentum
variables to reduce the problem to a single independent variable. The only
remaining difficulty is to handle the angular momentum term in the radial equation.

We end the Section with some questions:

\begin{enumerate}
\item To what degree can regularizations be accomplished by canonical
transformations? What is the relationship between regularizations and
canonical transformations?

\item What can be proved about extrema, boundedness and stability of orbits in
monotonic central potentials bounded by various power law potentials? in
monotonic central potentials? in arbitrary central potentials?
\end{enumerate}

\section{Inequivalent Lagrangians and Hamiltonians}

One of the more startling influences of quantum physics on the study of
classical mechanics is the realization that there exist inequivalent
Lagrangians determining a given set of classical paths. Inequivalent
Lagrangians for a given problem are those whose difference is \textit{not} a
total derivative. While it is not too surprising that a given set of paths
provides extremals for more than one functional, it is striking that some
systems permit infinitely many Lagrangians for the same paths. There remain
many open questions on this subject, with most of the results holding in only
one dimension.

The existence of classically inequivalent Hamiltonians is not so clear, since
there are far more transformations preserving Hamiltonian structure than there
are preserving Lagrangian structure. However, distinct Hamiltonians abound in
quantum theory, where equivalent Hamiltonians may lead to distinct quantum
structures \cite{Balach.}. If there is more than one Hamiltonian for a system,
which one do we quantize? Furthermore, while it is clear that distinct
Hamiltonians can lead to different quantum theories, what about the converse?
Do there exist distinct Hamiltonian operators, $\hat{H},\hat{H}^{\prime}$ with
identical expectation values for all observables? Can distinct Hamiltonian
operators have the same energy spectra?

Here, we restrict our attention to classical questions. To begin our
exploration of inequivalent Lagrangians, we describe classes of free particle
Lagrangians and give some examples. Next we move to the theorems for $1$-dim
systems due to Yan, Kobussen and Leubner (\cite{Kobussen}, \cite{Yan1978},
\cite{Yan1981}, \cite{LeubnerZoller}, \cite{Leubner}) including a simple
example. Then we consider inequivalent Lagrangians in higher dimensions.
Finally, we briefly examine the possibilities for inequivalent Hamiltonians.

\subsection{General free particle Lagrangians}

There are distinct classes of Lagrangian even for free particle motion. We
derive the classes and give an example of each, noting how Galilean invariance
singles out the usual choice of Lagrangian.

The most general velocity dependent free particle Lagrangian is
\[
S=\int f(v)dt
\]
We assume the Cartesian form of the Euclidean metric, so that $v=\sqrt
{\delta_{ij}v^{i}v^{j}}.$ The equation of motion is
\[
\frac{d}{dt}\frac{\partial f}{\partial v^{i}}=0
\]
so the conjugate momentum
\[
p_{i}=\frac{\partial f}{\partial v^{i}}=f^{\prime}\frac{v_{i}}{v}%
\]
is conserved. We need only solve this equation for the velocity. Separating
the magnitude and direction, we have
\begin{align*}
\frac{v_{i}}{v}  &  =\frac{p_{i}}{p}\\
v  &  =g\left(  p\right)  \equiv\left[  f^{\prime}\right]  ^{-1}\left(
p\right)
\end{align*}
This solution is well-defined on any region in which the mapping between
velocity and momentum is $1-1.$ This means that velocity ranges may be any of
four types: $v\in\left(  0,\infty\right)  ,\left(  0,v_{1}\right)  ,\left(
v_{1},v_{2}\right)  ,\left(  v_{1},\infty\right)  .$ Which of the four types
occurs depends on the singularities of $f^{\prime}v^{i}/v.$ Since $v^{i}/v$ is
a well-defined unit vector for all nonzero $v_{i},$ it is $f^{\prime}$ which
determines the range. Requiring the map from $v\,_{i}$ to $p_{i}$ to be single
valued and finite, we restrict to regions of $f^{\prime}$ which are monotonic.
Independent physical ranges of velocity will then be determined by each zero
or pole of $f^{\prime}.$ In general there will be $n+1$ such ranges
\[
v\in(0,v_{1}),\left(  v_{1},v_{2}\right)  ,\ldots,\left(  v_{n},\infty\right)
\]
if there are $n$ singular points of $f^{\prime}$. Of course it is possible
that $v_{1}=0$ (so that on the lowest range, $\left(  0,v_{2}\right)  ,$ zero
velocity is forbidden), or $v_{1}=\infty$ so that the full range of velocities
is allowed. Within any of these regions, the Hamiltonian formulation is
well-defined and gives the same equations of motion as the Lagrangian formulation.

Thus, the motion for general $f$ may be described as follows. Picture the
space of all velocities divided into a number of spheres centered on the
origin. The radii of these spheres are given by the roots and poles of
$f^{\prime}.$ Between any pair of spheres, momentum and velocity are in $1-1$
correspondence and the motion is uniquely determined by the initial
conditions. In these regions the velocity remains constant and the resulting
motion is in a straight line. On spheres corresponding to zeros of $f^{\prime
}$, the direction of motion is not determined by the equation of motion. On
spheres corresponding to poles of $f^{\prime},$ no solutions exist. It is
amusing to note that all three cases occur in practice. We now give an example
of each.

First, consider the regular situation when $f^{\prime}$ is monotonic
everywhere so the motion is uniquely determined to be straight lines for all
possible initial velocities. The condition singles out the case of
unconstrained Newtonian mechanics. this is the only case that is Galilean
invariant, since Galilean boosts require the full range of velocities,
$v\in(0,\infty).$

When $f^{\prime}$ has zeros, we have situations where a complete set of
initial conditions is insufficient to determine the motion. Such a situation
occurs in Lovelock, or extended, gravity, in which the action in
$d$-dimensions (for $d$ even) is of the general form
\[
S=\sum_{k=0}^{d/2}a_{k}\int\mathbf{R}^{ab}\wedge\mathbf{R}^{cd}\wedge
\cdots\wedge\mathbf{R}^{ef}\wedge\mathbf{e}^{g}\wedge\cdots\wedge
\mathbf{e}^{h}\varepsilon_{abcd\cdots efg\cdots h}%
\]
where $\mathbf{R}^{ab}$ is the curvature $2$-form, $\mathbf{e}^{a}$ the solder
form and the $a_{k}$ are arbitrary constants. This is the most general curved
spacetime gravity theory in which the field equations depend on no higher than
second derivatives of the metric \cite{Lovelock}. In general, the field
equations depend on powers of the second derivatives of the metric, whereas in
general relativity this dependence is linear. Among the solutions are certain
special cases called ``geometrically free'' \cite{Wheeler86}. These arise as
follows. For some choices of the constants $a_{k},$ we may rewrite $S$ in the
form
\[
S=\int\prod_{k=0}^{d/2}\left(  \mathbf{R}^{a_{k}b_{k}}-\alpha_{k}%
\mathbf{e}^{a_{k}}\mathbf{e}^{b_{k}}\right)  \varepsilon_{a_{1}b_{1}\cdots
a_{d/2}b_{d/2}}%
\]
Suppose that for all $k=1,\ldots,n$ for some $n$ in the range $2<n<d/2,$ we
have
\[
\alpha_{k}=\alpha
\]
for some fixed value $\alpha.$ Then the variational equations all contain at
least $n-1$ factors of
\[
\mathbf{R}^{a_{k}b_{k}}-\alpha\mathbf{e}^{a_{k}}\mathbf{e}^{b_{k}}%
\]
Therefore, if there is a subspace of dimension $m>d-n+1$ of constant
curvature
\[
\mathbf{R}^{ab}=\alpha\mathbf{e}^{a}\mathbf{e}^{b}%
\]
for $a,b=1,\ldots,m,$ then the field equations are satisfied regardless of the
metric on the complementary subspace. This is similar to the case of vanishing
$f^{\prime},$ where the equation of motion is satisfied regardless of the
direction of the velocity,
\[
p_{i}=f^{\prime}\frac{v_{i}}{v}\equiv0
\]
as long as $v,$ but not $v_{i}$, is constant.

Finally, suppose $f^{\prime}$ has a pole at some value $v_{0}.$ Then the
momentum diverges and motion never occurs at velocity $v_{0}.$ Of course, this
is the case in special relativity, where the action of a free particle may be
written as
\begin{align*}
S &  =\int p_{\alpha}dx^{\alpha}=-\int Edt+p_{i}dx^{i}\\
&  =-mc^{2}\int\sqrt{1-\frac{v^{2}}{c^{2}}}\ dt.
\end{align*}
With $f(v)=-mc^{2}\sqrt{1-v^{2}/c^{2}},$ we have
\[
f^{\prime}=\frac{mv}{\sqrt{1-v^{2}/c^{2}}}%
\]
with the well known pole in momentum at $v=c.$

Note that there is a complementary situation for Hamiltonians. From the
Lagrangians for the free particle,
\[
S=\int f(v)dt
\]
we have the conjugate momenta
\[
p_{i}=f^{\prime}\frac{v_{i}}{v},\text{ \ }p=f^{\prime}=g
\]
and Hamiltonians
\[
H=vf^{\prime}-f.
\]
Hamilton's equations are
\begin{align*}
\frac{dx^{i}}{dt}  &  =\frac{\partial H}{\partial p_{i}}=\frac{p^{i}}{p}%
g^{-1}\left(  p\right) \\
\frac{dp_{i}}{dt}  &  =-\frac{\partial H}{\partial x^{i}}=0
\end{align*}
Once again, the constancy of the momentum is immediate. However, despite the
apparent diversity of Hamiltonians, they are locally related by canonical
transformations. The \textit{only} distinctions are the global ones, and these
exactly match those described above.

\subsection{Inequivalent Lagrangians}

The existence of inequivalent Lagrangians for a given physical problem seems
to trace back to Lie \cite{Lie}. Dirac (\cite{Dirac1933},\cite{Dirac1950}) was
certainly well aware of the ambiguities involved in passing between the
Lagrangian and Hamiltonian formulations of classical mechanics. Later, others
(\cite{Whittaker},\cite{Wolsky},\cite{CurieSaletan},\cite{Currie}), identified
certain non-canonical transformations which nonetheless preserve certain
Hamiltonians. A specific non-canonical transformation of the $2$-dim harmonic
oscillator is provided by Gelman and Saletan \cite{GelmanSaletan}. Bolza
\cite{Bolza} showed that independent Lagrangians can give the same equations
of motion and, a few years later, Kobussen \cite{Kobussen}, Yan
(\cite{Yan1978},\cite{Yan1981}) and Okubo (\cite{Okubo1},\cite{Okubo2})
independently gave systematic developments showing that an infinite number of
inequivalent Lagrangians exist for $2$-dim mechanical systems. Shortly
thereafter, Leubner \cite{Leubner} generalized and streamlined Yan's proof to
include arbitrary functions of two constants of motion.

Leubner's result, the most general to date, may be stated as follows. Given
any two constants of motion, $\left(  \alpha,\beta\right)  ,$ associated with
the solution to a given $1$-dim equation of motion, the solution set for any
Lagrangian of the form
\begin{align}
L\left(  x,\dot{x},t\right)   &  =\int_{v}^{\dot{x}}\frac{\dot{x}-v}{v}\left|
\frac{\partial\left(  \alpha,\beta\right)  }{\partial\left(  v,t\right)
}\right|  dv\nonumber\\
&  +\int_{x_{0}}^{x}f\left(  \tilde{x},v_{0},t\right)  \frac{1}{v_{0}}\left|
\frac{\partial\left(  \alpha,\beta\right)  }{\partial\left(  v,t\right)
}\right|  _{v=v_{0}}d\tilde{x}+\frac{d\Omega}{dt} \label{Leubner theorem}%
\end{align}
where $\left|  \frac{\partial\left(  \alpha,\beta\right)  }{\partial\left(
v,t\right)  }\right|  $ is the Jacobian, includes the same solutions locally.
Notice that $\alpha$ and $\beta$ are arbitrary constants of the motion -- each
may be an arbitrary function of simpler constants such as the Hamiltonian. We
argue below that in $1$-dim the solution sets are locally identical, though
\cite{Leubner} provides no explicit proof. In higher dimensions there are easy counterexamples.

We illustrate a special case of this formula, of the form
\begin{equation}
L\left(  x,v\right)  =\dot{x}\int^{\dot{x}}\frac{K\left(  x,v\right)  }{v^{2}%
}dv \label{Alternate Lagrangians}%
\end{equation}
where $K$ is any constant of the motion of the system. This expression is
valid when the original Lagrangian has no explicit time dependence. Following
Okubo \cite{Okubo2}, we prove that eq.(\ref{Alternate Lagrangians}) leads to
the constancy of $K.$ The result follows immediately from the Euler-Lagrange
expression for $L:$%
\begin{align*}
\frac{d}{dt}\frac{\partial L}{\partial\dot{x}}-\frac{\partial L}{\partial x}
&  =\frac{d}{dt}\left(  \int^{\dot{x}}\frac{K\left(  x,v\right)  }{v^{2}%
}dv+\dot{x}\frac{K\left(  x,\dot{x}\right)  }{\dot{x}^{2}}\right)  -\dot
{x}\int^{\dot{x}}\frac{1}{v^{2}}\frac{\partial K\left(  x,v\right)  }{\partial
x}dv\\
&  =\frac{\ddot{x}}{\dot{x}}\frac{\partial K\left(  x,\dot{x}\right)
}{\partial\dot{x}}+\frac{\partial K\left(  x,\dot{x}\right)  }{\partial x}\\
&  =\frac{1}{\dot{x}}\frac{dK\left(  x,\dot{x}\right)  }{dt}.
\end{align*}
Therefore, the Euler-Lagrange equation holds if and only if $K\left(
x,\dot{x}\right)  $ is a constant of the motion.

The uniqueness in $1$-dim follows from the fact that a single constant of the
motion is sufficient to determine the solution curves up to the initial point.
The uniqueness also depends on there being only a single Euler-Lagrange
equation. These observations lead us to a higher dimensional result below.

It is interesting to notice that we can derive this form for $L,$ but with $K$
replaced by the Hamiltonian, by inverting the usual expression,
\[
H=\dot{x}\frac{\partial L}{\partial\dot{x}}-L
\]
for the Hamiltonian in terms of the Lagrangian. First, rewrite the right side
as:
\[
H=\dot{x}\frac{\partial L}{\partial\dot{x}}-L=\dot{x}^{2}\frac{\partial
}{\partial\dot{x}}\left(  \frac{L}{\dot{x}}\right)  .
\]
Now, dividing by $\dot{x}$ and integrating (regarding $H$ as a function of the
velocity) we find:
\[
L=\dot{x}\int^{\dot{x}}\frac{H\left(  x,v\right)  }{v^{2}}dv
\]
The remarkable fact is that the Hamiltonian may be replaced by any constant of
the motion in this expression. Conversely, suppose we begin with the
Lagrangian in terms of an arbitrary constant of motion, $K,$ according to
eq.(\ref{Alternate Lagrangians}),
\[
L\left(  x,v\right)  =\dot{x}\int^{\dot{x}}\frac{K\left(  x,v\right)  }{v^{2}%
}dv
\]
Then constructing the conserved Hamiltonian,
\begin{align*}
\tilde{H}\left(  x,p\right)   &  =\dot{x}\frac{\partial L}{\partial\dot{x}%
}-L\\
&  =\dot{x}\frac{\partial}{\partial\dot{x}}\left(  \dot{x}\int^{\dot{x}}%
\frac{K\left(  x,v\right)  }{v^{2}}dv\right)  -\dot{x}\int^{\dot{x}}%
\frac{K\left(  x,v\right)  }{v^{2}}dv\\
&  =\dot{x}\left(  \int^{\dot{x}}\frac{K\left(  x,v\right)  }{v^{2}}%
dv+\frac{K\left(  x,\dot{x}\right)  }{\dot{x}}\right)  -\dot{x}\int^{\dot{x}%
}\frac{K\left(  x,v\right)  }{v^{2}}dv\\
&  =K\left(  x,\dot{x}\right)
\end{align*}
we arrive at the chosen constant of motion! This proves the
Gelman-Saletan-Currie conjecture \cite{GelmanSaletan}: any nontrivial
time-independent constant of motion gives rise to a possible Hamiltonian.
Proofs of the conjecture are due to Yan (\cite{Yan1978},\cite{Yan1981}) and
Leubner \cite{Leubner}.

The conjugate momentum to $L$ constructed according to
eq.(\ref{Alternate Lagrangians}) is
\begin{align*}
\tilde{p}  &  =\frac{\partial L}{\partial\dot{x}}\\
&  =\frac{\partial}{\partial\dot{x}}\left(  \dot{x}\int^{\dot{x}}%
\frac{K\left(  x,v\right)  }{v^{2}}dv\right) \\
&  =\int^{\dot{x}}\frac{K\left(  x,v\right)  }{v^{2}}dv+\frac{K\left(
x,\dot{x} \right)  }{\dot{x}}%
\end{align*}
Of course, if $K=\frac{1}{2}m\dot{x}^{2}+V,$ both $\tilde{H}$ and $\tilde{p}$
reduce to the usual expressions.

The simple harmonic oscillator is sufficient to illustrate the method
(\cite{Lopez},\cite{GhoshCJP}). Since the Hamiltonian, $H=\frac{1}{2}%
mv^{2}+\frac{1}{2}kx^{2},$ is a constant of the motion so is $H^{2},$ so we
write
\begin{align*}
L  &  =\frac{1}{4}\dot{x}\int^{\dot{x}}\frac{1}{v^{2}}\left(  m^{2}%
v^{4}+2kmv^{2}x^{2}+k^{2}x^{4}\right)  dv\\
&  =\frac{1}{12}m^{2}\dot{x}^{4}+\frac{1}{2}km\dot{x}^{2}x^{2}-\frac{1}%
{4}k^{2}x^{4}.
\end{align*}
The Euler-Lagrange equation resulting from $L$ is
\begin{align*}
0  &  =\frac{d}{dt}\frac{\partial L}{\partial\dot{x}}-\frac{\partial
L}{\partial x}\\
&  =\frac{d}{dt}\left(  \frac{1}{3}m^{2}\dot{x}^{3}+km\dot{x}x^{2}\right)
-\left(  km\dot{x}^{2}x-k^{2}x^{3}\right) \\
&  =\left(  m\ddot{x}+kx\right)  \left(  m\dot{x}^{2}+kx^{2}\right)  .
\end{align*}
Either of the two factors may be zero. Setting the first to zero is gives the
usual equation for the oscillator, while setting the second to zero we find
the same solutions in exponential form:
\[
x=Ae^{i\omega t}+Be^{-i\omega t}%
\]

The conjugate momentum and Hamiltonian are:
\begin{align*}
\tilde{H}\left(  x,\tilde{p}\right)   &  =H^{2}\left(  x,p\right)  =\frac
{1}{4}\left(  m^{2}\dot{x}^{4}+2km\dot{x}^{2}x^{2}+k^{2}x^{4}\right)  \\
\tilde{p} &  =\frac{\partial L}{\partial\dot{x}}=\frac{1}{3}m^{2}\dot{x}%
^{3}+km\dot{x}x^{2}.
\end{align*}
While it is possible to solve the cubic equation to find $\dot{x}\left(
\tilde{p}\right)  ,$ and then substitute to find $\tilde{H}\left(  x,\tilde
{p}\right)  $ as an explicit function of $\tilde{p},$ it is clear that the
resulting expression is not of the same form as the original Hamiltonian.
There remains the question of whether this effect could be achieved by a
time-independent canonical transformation. The transformation of the
momentum,
\[
\tilde{p}=\frac{p^{3}}{3m}+kpx^{2}%
\]
is part of a canonical transformation, given by
\begin{align*}
\tilde{x} &  =-\frac{1}{2k}\ln p\\
\tilde{p} &  =\frac{p^{3}}{3m}+kpx^{2}%
\end{align*}
However, this does not simplify the form of the Hamiltonian. We find:
\[
\tilde{H}\left(  x,\tilde{p}\right)  =\frac{1}{4m^{2}}\left(  \frac{4}%
{9}e^{-8k\tilde{x}}+m^{2}\tilde{p}^{2}e^{4k\tilde{x}}+\frac{4}{3}m\tilde
{p}e^{-2k\tilde{x}}\right)
\]
and the resulting Hamiltonian equations of motion are not transparent.

There does exist, of course, a time-dependent canonical transformation
relating the two Hamiltonians. The systems are nonetheless distinct globally,
since the cubic relationship between momentum and velocity limits the allowed
ranges of the variables for the higher order Hamiltonian. It would be
interesting to know if there is a unique Hamiltonian for which $p\left(
v\right)  $ is $1-1.$

\subsubsection{Are inequivalent Lagrangians equivalent?}

Inequivalent Lagrangians have been defined as Lagrangians which lead to the
same equations of motion but differ by more than a total derivative. For the
simple case above, the cubic order equation of motion factors into the energy
times the usual equation of motion, and setting either factor to zero gives
the usual solution and only the usual solution. However, is this true in
general? The Yan-Leubner proof shows that the new Lagrangian has the same
solutions, but how do we know that none of the higher order Lagrangians
introduces spurious solutions? The proofs do not address this question
explicitly. If some of these Lagrangians introduce extra solutions, then they
are not really describing the same motions.

Suppose, for some time-independent Hamiltonian we write
\[
L=v\int^{v}\frac{f[\alpha\left(  x,\xi\right)  ]}{\xi^{2}}d\xi
\]
where $\alpha$ is any constant of the motion. Then we know that the
Euler-Lagrange equation is satisfied by the usual equation of motion. But what
\textit{is} the Euler-Lagrange equation? We have shown that
\[
\frac{d}{dt}\frac{\partial L}{\partial\dot{x}}-\frac{\partial L}{\partial
x}=\frac{1}{\dot{x}}\frac{dK\left(  x,\dot{x}\right)  }{dt}=\frac{1}{\dot{x}%
}f^{\prime}\frac{d\alpha\left(  x,\dot{x}\right)  }{dt}.
\]
Setting this to zero, we have two types of solution
\begin{align*}
f^{\prime}\left(  \alpha\right)   &  =0\\
\frac{d\alpha}{dt} &  =0.
\end{align*}
If spurious solutions could arise from motions with $f^{\prime}=0$, those
motions would have to stay at the critical point, $\alpha_{0}$ say, of $f.$
But this means that $\alpha=\alpha_{0}$ remains constant. Therefore, the only
way to introduce spurious solutions is if $d\alpha/dt=0$ has solutions beyond
the usual solutions. This may not be possible in one dimension. Finally, the
inverse of the equation $\alpha\left(  x,t\right)  =\alpha_{0}$ may not exist
at critical points, so the theorem must refer only to \textit{local}
equivalence of the solutions for inequivalent Lagrangians.

\subsection{Inequivalent Lagrangians in higher dimensions}

It is of interest to extend the results on inequivalent systems to higher
dimension. Presumably, the theorems generalize in some way, but while one
dimensional problems may be preferable ``for simplicity'' \cite{Leubner}, this
restricted case has many special properties that may not generalize. In any
case, the method of proof of the Kobussen-Yan-Leubner theorem does not
immediately generalize.

For $1$-dim classical mechanics, there are only two independent constants of
motion. The Kobussen-Yan-Leubner theorem, eq.(\ref{Leubner theorem}), makes
use of one or both to characterize the Lagrangian and, as noted above, one
constant can completely determine the paths motion in $1$-dim. The remaining
constant is required only to specify the initial point of the motion. This
leads to a simple conjecture for higher dimensions, namely, that the paths are
in general determined by $n$ of the $2n$ constants of motion. This is because
$n$ of the constants specify the initial position, while the remaining
constants determine the paths.

We make these comments concrete with two examples. First, consider again the
free particle in $n$-dim. The usual Hamiltonian is
\[
H=\frac{\mathbf{p}^{2}}{2m}%
\]
and we immediately find that a complete solution is characterized by the
initial components of the momentum, $p_{0i}$ and the initial position,
$x_{0i}.$ Clearly, knowledge of the momenta is necessary and sufficient to
determine a set of flows. If we consider inequivalent Lagrangians
\[
L=v\int^{v}\frac{f\left(  \xi\right)  }{\xi^{2}}d\xi=F\left(  v\right)
\]
where
\[
v=\sqrt{\mathbf{v}^{2}}%
\]
then the momenta
\[
p_{i0}=\frac{\partial L}{\partial v^{i}}=F^{\prime}\frac{v_{i}}{v}%
\]
comprise a set of first integrals of the motion. Inverting for the velocity
\[
v^{i}=v^{i}\left(  p_{i0}\right)
\]
fixes the flow without fixing the initial point.

In general we will need at least this same set of relations, $v^{i}%
=v^{i}\left(  p_{i0}\right)  ,$ to determine the flow, though the generic case
will involve $n$ relations depending on $2n$ constants:
\[
v^{i}=v^{i}\left(  p_{i0},x_{0}^{i}\right)  .
\]
Notice that fewer relations does not determine the flow even for free motion
in two dimensions. Thus, knowing only
\[
v_{x}=\frac{p_{0x}}{m}%
\]
leaves the motion in the $y$ direction fully arbitrary.

In an arbitrary number of dimensions, we find that expression for the energy
in terms of the Lagrangian is still integrable as in the $1$-dim case above,
as long as $v=\sqrt{\mathbf{v}^{2}}.$ If the Lagrangian does not depend
explicitly on time, then energy is conserved. Then, letting $\hat{\theta}%
^{i}=\dot{x}^{i}/v,$ we can still write the Lagrangian as an integral over
Hamiltonian:
\[
L\left(  \mathbf{x},v,\hat{\theta}_{v}\right)  =v\int^{v}\frac{H\left(
x,\xi,\hat{\theta}\right)  }{\xi^{2}}d\xi+f\left(  \mathbf{x},\hat{\theta}%
_{v}\right)
\]
where $f\left(  \mathbf{x},\vec{\theta}_{v}\right)  $ is now necessary in
order for $L$ to satisfy the Euler-Lagrange equations. The integral term of
this expression satisfies \textit{one} of the Euler-Lagrange equations. If we
now define a new Lagrangian by replacing $H$ by an arbitrary, time-independent
constant of the motion, $\alpha\left(  x,v,\hat{\theta}\right)  ,$%
\[
\tilde{L}=v\int^{v}\frac{\alpha\left(  x,\xi,\hat{\theta}\right)  }{\xi^{2}%
}d\xi+f\left(  x,\hat{\theta}\right)
\]
then the new Lagrangian, $\tilde{L},$ still satisfies the same Euler-Lagrange
equation
\[
\dot{x}^{i}\left(  \frac{d}{dt}\frac{\partial\tilde{L}}{\partial\dot{x}^{i}%
}-\frac{\partial\tilde{L}}{\partial x^{i}}\right)  =0.
\]
We conjecture that for a suitable choice of $f,$ $\tilde{L}$ provides an
inequivalent Lagrangian, thereby providing \textit{one} of the $n$ relations
required to specify the flow.

\subsection{Inequivalent Hamiltonians}

The question of inequivalent Hamiltonians is quite distinct from that of
inequivalent Lagrangians, because the group of allowed transformations is much
larger. Indeed, Hamilton-Jacobi theory shows that any Hamiltonian may be made
trivial by a canonical transformation. This also means that any two
Hamiltonians are related locally by a time-dependent canonical transformation.
At least in this sense, all Hamiltonians with the same number of degrees of
freeedom are locally equivalent.

In light of this, the first question to be answered is the following. Since
the properties of canonical transformations are defined by the demand that
they change the action by no more than a total derivative (for example, in the
derivation of the properties of generating functions), how can two
Hamiltonians be locally equivalent while the corresponding Lagrangians are
inequivalent? The answer is subtle. Canonical transformations are defined in
such a way as to leave the Hilbert one-form
\[
L\left(  p_{i},\dot{x}^{i},q^{i},t\right)  dt=p_{i}\frac{dx^{i}}%
{dt}dt-H\left(  p_{j},q^{k}\right)  dt
\]
changed by no more than an exact form. But this $L$ is not quite the
Lagrangian, since
\[
L=L\left(  p_{i},\dot{x}^{i},q^{i},t\right)
\]
while a true Lagrangian is a function of $x^{i}$ and $\dot{x}^{i}$ only. Since
the Lagrangian formalism is invariant only under coordinate diffeomorphisms,
$x^{i}=x^{i}\left(  q^{j},t\right)  ,$ canonical transformations involving
both $p_{i}$ and $x^{i}$ are not expected to preserve it.

Despite Hamilton-Jacobi theory, there are ways to define a notion of
inequivalent Hamiltonians. First, if we restrict to time-independent canonical
transformations, there exist distinct Hamiltonian systems related by
diffeomorphisms, $H\left(  q,\pi\right)  =H[x\left(  q,\pi\right)  ,p\left(
q,\pi\right)  ]$. This allows substantial variation in the functional form of
the Hamiltonian and it may be difficult to determine whether two Hamiltonians
are related in this way. Second, we may define canonoid transformations,
defined as preserving the canonical structure for one or more Hamilton
\cite{GelmanSaletan}. Presumably, these are related to symmetries of
particular systems. Or, third, we may quantize the system and ask whether the
quantum systems are equivalent.

These considerations point to a heirarchy of classifications of Hamiltonian
systems, equivalent up to some set of transformations. It would be useful to
know the exact set of transformations under which a given set of phase space
paths is invariant. We know that the set is smaller than time-dependent
transformations and larger than canonical transformations, and seems likely
that the answer depends on the class of curves in some way. Since it is really
the solution curves that define equivalence, it is clear that some systems
will have more symmetry than others.

Even if we had a clear characterization of Hamiltonians for a given system, it
is not clear that we know what the system is. For example, suppose a given
system admits a time-independent canonical tranformation taking the
Hamiltonian to a constant. Such a system exists if we can find nontrivial
solutions to the time-independent Hamilton-Jacobi equation
\[
\frac{m}{2}\vec{\triangledown}S\cdot\vec{\triangledown}S+V=0
\]
and this is surely possible for some systems. But this means that one of the
equivalent formulations of the problem describes straight-line motion.
Clearly, finding the solution curves is not enough to describe such a system
-- we must also keep track of the sequence of transformations we used to
trivialize that solution. Thus, specifying a classical system requires some
statement about the correspondence between phase space coordinates and
measurements in some physical system. Of course, there are many equivalent
ways to set up such a correspondence, but at least one must be specified and
the subsequent transformations tracked.

\section{Constants of motion}

Recent decades have seen interesting new techniques and revivals of known
results for symmetries (\cite{Vujanovic},\cite{GorringeLeach}). Some of these
have to do with the Kepler problem. The best-known rediscovery concerning the
Kepler problem is that in addition to the energy, $E$ and angular momentum,
\begin{align}
E  &  =\frac{1}{2}m\mathbf{\dot{x}}^{2}-\frac{\alpha}{r}\nonumber\\
\mathbf{L}  &  =\mathbf{r}\times\mathbf{p} \label{E and L}%
\end{align}
the Laplace-Runge-Lenz vector (\cite{Laplace}, \cite{Runge}, \cite{Lenz},
\cite{GoldsteinMore}) is conserved. We define the Laplace-Runge-Lenz vector
by
\[
\mathbf{A}=\mathbf{p}\times\mathbf{L}-m\alpha\mathbf{\hat{r}.}%
\]
Geometrically, $\mathbf{A}$ points in the direction of the periapsis, and may
therefore be thought of as giving specifying the orientation of the orbit
within the orbital plane. Keplerian orbits can be described completely in
terms of six initial conditions, and since one of these is the initial
position on a given ellipse, only five remain among the energy, angular
momentum and Laplace-Runge-Lenz vector \cite{GoldsteinPS}. Two constraints --
the orthogonality of $\mathbf{A}$ and $\mathbf{L,}$ and a relationship between
the magnitudes $A,L$ and $E$ -- give the correct count. Of course, these three
quantities are not the only set of constants we can choose. A number of fairly
recent authors (\cite{Munoz},\cite{AbelsondiSessaRudolph},\cite{Patera}%
,\cite{Derbes}) have identified a simpler conserved vector quantity, which
(lacking evidence for an earlier reference) we will call the Hamilton vector
\cite{Hamilton}. It is given by
\[
\mathbf{u}=\mathbf{v}-\frac{\alpha}{L}\mathbf{\hat{\varphi}}%
\]
and may be used together with the energy and angular momentum as a complete
set of constants. Apparently this vector was well-known in the 19th century,
then dropped from texts \cite{Munoz}. Its time constancy is a direct
consequence of the force law since, for an arbitrary central force $f(r),$
\begin{align}
\frac{du_{i}}{dt}  &  =\frac{dv_{i}}{dt}-\frac{\alpha}{L}\frac{d\hat{\varphi
}_{i}}{dt}=-\frac{f\left(  r\right)  x_{i}}{mr}+\frac{\alpha}{mr^{2}%
\dot{\varphi}}\frac{d\varphi}{dt}\frac{x_{i}}{r}\\
&  =\left[  \frac{\alpha}{r^{2}}-f\left(  r\right)  \right]  \frac{x_{i}}{mr}%
\end{align}
and $du_{i}/dt=0$ precisely when $f(r)$ is given by an inverse square law .
Notice that it is the balance between the radial dependences of angular
momentum and the force that allow this characterization. We give a derivation
of the Hamilton vector below.

The Laplace-Runge-Lenz vector, the Hamilton vector and the angular momentum
are related by
\[
\mathbf{A}=m\mathbf{u}\times\mathbf{L}%
\]
and
\begin{align}
\mathbf{L}\times\mathbf{A}  &  =m\mathbf{L}\times\left(  \mathbf{u}%
\times\mathbf{L}\right) \\
&  =mL^{2}\mathbf{u}%
\end{align}
where we have used the fact that $\mathbf{u}$ lies in the plane of the orbit
and is therefore perpendicular to $\mathbf{L}$.

It might be of interest to study more general central force problems using
time-dependent versions of the Laplace-Runge-Lenz and Hamiltonian vectors.
While no longer conserved, they are more geometrical than the usual polar
coordinates. This could result in some simplification. It would also be of
interest to know if these vectors correspond to symmetries -- perhaps they
reflect symmetries of the corresponding phase space, or, since the bound state
Kepler problem may be embedded with an $SO(4)$ symmetry, perhaps they are part
of that symmetry. It seems likely that the Kepler problem has an even larger
symmetry -- perhaps $SO(4,1)$ -- since the open orbits have $SO(3,1)$
symmetry. This might be explored by writing the Kepler problem in terms of
$Spin\left(  4,1\right)  $ conformal spinors.

Mu\~{n}oz \cite{Munoz} shows that the Hamiltion vector leads to an easy
derivation of the equation of motion. Indeed, let the perihelion of the orbit
occur at time $t=0$ on the $x$-axis so that the velocity is given by
\[
\mathbf{v}=v_{0}\mathbf{\hat{\varphi}}%
\]
Then $\mathbf{u}=u\mathbf{\hat{\jmath}}$, where the unit vector in the
y-direction gives the initial direction of $\mathbf{\hat{\varphi}.}$ Dotting
$\mathbf{u}$ with $\mathbf{\hat{\varphi}}$ we have
\begin{align*}
\mathbf{u\cdot\hat{\varphi}}  &  =\mathbf{v\cdot\hat{\varphi}}-\alpha/L\\
u\cos\phi &  =r\dot{\varphi}-\alpha/L
\end{align*}
or replacing $\dot{\varphi}=L/(mr^{2}),$
\begin{align}
\frac{1}{r}  &  =\frac{mu}{L}\cos\varphi+\frac{\alpha m}{L^{2}}\nonumber\\
r  &  =\frac{L^{2}/m\alpha}{1+(Lu_{0}/\alpha)\cos\varphi}
\label{Kepler orbits}%
\end{align}
as usual.

Tjiang and Sutanto \cite{TjiangSutanto} describe a straightforward way to
identify constants of the motion arising from the vanishing of
\begin{equation}
\frac{df}{dt}=\left[  f,H\right]  +\frac{\partial f}{\partial t}
\label{Const of motion}%
\end{equation}
based on a well-known theorem on the solution of differential equations. The
theorem states that any equation of the form
\[
\sum_{i}P_{i}\left(  x_{1},\ldots,x_{n}\right)  \frac{\partial f}{\partial
x_{i}}=R\left(  x_{1},\ldots,x_{n},f\right)
\]
has the general solution given by
\[
f=\Phi\left(  u_{1},\ldots,u_{k}\right)
\]
where $k\leq n$ and the $u_{i}\left(  x_{1},\ldots,x_{n},f\right)  $ are
solutions to
\[
\frac{dx_{1}}{P_{1}}=\frac{dx_{2}}{P_{2}}=\cdots=\frac{dx_{n}}{P_{n}}%
=\frac{df}{R}%
\]
Applied to eq.(\ref{Const of motion}), the theorem implies that the functions
$u_{i}$ are all constants of the motion. Moreover, we can compute the possible
constants of motion by solving the equations
\[
\frac{dx_{1}}{\left(  \frac{\partial H}{\partial p_{1}}\right)  }=\cdots
=\frac{dx_{n}}{\left(  \frac{\partial H}{\partial p_{n}}\right)  }%
=\frac{dp_{1}}{\left(  -\frac{\partial H}{\partial x_{1}}\right)  }%
=\cdots=\frac{dx_{n}}{\left(  -\frac{\partial H}{\partial x_{n}}\right)  }=dt
\]

To illustrate the method and at the same time derive the Hamilton vector as a
constant of Keplerian motion, we apply the technique to the Kepler problem.
First, the Hamiltonian is given, in any dimension $n\geq2,$ by
\[
H=\frac{\mathbf{p}^{2}}{2m}-\frac{\alpha}{r}%
\]
so we must solve the equations
\begin{equation}
\frac{m}{p_{1}}dx_{1}=\cdots=\frac{m}{p_{n}}dx_{n}=-\frac{r^{3}}{\alpha x_{1}%
}dp_{1}=\cdots=-\frac{r^{3}}{\alpha x_{n}}dp_{n}=dt \label{Tjiang-Sutanto eqs}%
\end{equation}
First, for each $i,$ consider the equations of the form
\begin{align*}
\frac{m}{p_{i}}dx_{i}  &  =-\frac{r^{3}}{\alpha x_{i}}dp_{i}\\
\frac{\alpha x_{i}}{r^{3}}dx_{i}+\frac{1}{m}p_{i}dp_{i}  &  =0.
\end{align*}
Summing over $i$ we have the constancy of the Hamiltonian:
\[
dH=d\left(  \sum\frac{p_{i}^{2}}{2m}-\frac{\alpha}{r}\right)  =0
\]
Next, consider the equations among the $dx_{i}.$ For any pair of these
$\left(  i\neq j\right)  $we have
\[
\frac{m}{p_{i}}dx_{i}=\frac{m}{p_{j}}dx_{j}%
\]
so that
\begin{align}
0  &  =p_{j}dx_{i}-p_{i}dx_{j}\\
&  =d\left(  p_{j}x_{i}-p_{i}x_{j}\right)  -\left(  dp_{j}x_{i}-dp_{i}%
x_{j}\right)
\end{align}
We may replace the momentum differentials using the Tjiang-Sutanto equations,
eqs.(\ref{Tjiang-Sutanto eqs}) to write
\[
dp_{j}x_{i}-dp_{i}x_{j}=-\frac{\alpha x_{j}}{r^{3}}\frac{mx_{i}}{p_{1}}%
dx_{1}+\frac{\alpha x_{i}}{r^{3}}\frac{mx_{j}}{p_{1}}dx_{1}=0
\]
so we have conservation of all of the components
\[
M_{ij}=x_{i}p_{j}-x_{j}p_{i}%
\]
of the angular momentum. Of course, in $3$-dimensions we may use the
Levi-Civita tensor to write this as
\[
L_{i}=\frac{1}{2}\varepsilon_{ijk}\left(  x_{i}p_{j}-x_{j}p_{i}\right)
=\left[  \mathbf{x}\times\mathbf{p}\right]  _{i}%
\]
In Sec. 2 we showed from the constancy of $M_{ij}$ that the motion remains in
a fixed plane for all time.

Finally, we study the additional constants of motion arising from equations of
the form
\[
\frac{m}{p_{1}}dx_{1}=-\frac{r^{3}}{\alpha x_{i}}dp_{i}%
\]
Since $p_{i}=mdx_{i}/dt,$ this may be written as
\begin{align}
dt+\frac{mr^{3}}{\alpha x_{i}}dv_{i}  &  =0\\
\frac{\alpha x_{i}}{mr^{3}}dt+dv_{i}  &  =0
\end{align}
Now, using $L=mr^{2}\dot{\varphi}$ where $L$ is the magnitude of the angular
momentum, $L=\sqrt{\sum M_{ij}M_{ij}},$ and $\dot{\varphi}$ is in the plane of
the orbit, we have
\begin{align}
dv_{1}+\frac{\alpha}{L}\frac{x_{1}}{r}d\varphi &  =0\\
d\left(  v_{1}+\frac{\alpha}{L}\sin\varphi\right)   &  =0
\end{align}
and similarly
\[
d\left(  v_{1}-\frac{\alpha}{L}\cos\varphi\right)  =0
\]
Adding these with unit vectors and noting that $\mathbf{v}=v_{1}%
\mathbf{\hat{\imath}}+v_{2}\mathbf{\hat{\jmath}}$ comprises the entire
velocity vector, establishes the constancy of the Hamilton vector:
\[
\mathbf{u}=\mathbf{v}-\frac{\alpha}{L}\mathbf{\hat{\varphi}}%
\]
Thus, $\left\{  H,M_{ij},u_{i}\right\}  $ is a complete set of constants of
the motion for the $n$-dim Kepler problem. The solution,
eq.(\ref{Kepler orbits}), follows immediately.

\section{Higher order equations of motion}

\subsection{Euler-Lagrange and Hamiltonian systems of arbitrary order}

There generalization of the Euler-Lagrange equation to systems for which the
Lagrangian depends on higher than second derivatives of the postion is
immediate and well-known. If $L=L\left(  x,\dot{x},\ddot{x},\ldots,x^{\left(
n\right)  },t\right)  $ the resulting variation leads to
\[
\sum_{k=0}^{n}\left(  -\right)  ^{k}\frac{d^{k}}{dt^{k}}\frac{\partial
L}{\partial x^{\left(  k\right)  }}=0
\]
This generalized Euler-Lagrange equation is generically of order $2n.$ Such
systems are, of course, allowed within even the Newtonian formulation. A
simple electronic circuit and feedback mechanism can easily drive a motor in a
way dependent upon rates of change of the acceleration. Moreover, there are
systems of genuine physical and mathematical interest which require higher
order differential equations for their description. After deriving a few
general results for higher order systems, we look at one example in detail --
Bohmian quantum mechanics.

Returning to generalised Euler-Lagrange systems, suppose $L$ is independent of
time. Then%

\[
\frac{dL}{dt}=\sum_{k=0}^{n}x^{\left(  k+1\right)  }\frac{\partial L}{\partial
x^{\left(  k\right)  }}%
\]
But
\begin{align}
x^{\left(  k+1\right)  }\frac{\partial L}{\partial x^{\left(  k\right)  }}  &
=\frac{d}{dt}\left(  x^{\left(  k\right)  }\frac{\partial L}{\partial
x^{\left(  k\right)  }}\right)  -x^{\left(  k\right)  }\frac{d}{dt}%
\frac{\partial L}{\partial x^{\left(  k\right)  }}\\
&  =\frac{d}{dt}\left(  x^{\left(  k\right)  }\frac{\partial L}{\partial
x^{\left(  k\right)  }}\right)  -\frac{d}{dt}\left(  x^{\left(  k-1\right)
}\frac{d}{dt}\frac{\partial L}{\partial x^{\left(  k\right)  }}\right) \\
&  +x^{\left(  k-1\right)  }\frac{d^{2}}{dt^{2}}\frac{\partial L}{\partial
x^{\left(  k\right)  }}\\
&  \vdots\\
&  =\sum_{m=0}^{k-1}\left(  -\right)  ^{m}\frac{d}{dt}\left(  x^{\left(
k-m\right)  }\frac{d^{m}}{dt^{m}}\frac{\partial L}{\partial x^{\left(
k\right)  }}\right)  -x^{\left(  1\right)  }\left(  -\right)  ^{k-1}%
\frac{d^{k}}{dt^{k}}\frac{\partial L}{\partial x^{\left(  k\right)  }}%
\end{align}
so
\[
\frac{dL}{dt}=\frac{d}{dt}\sum_{k=0}^{n}\sum_{m=0}^{k-1}\left(  -\right)
^{m}\left(  x^{\left(  k-m\right)  }\frac{d^{m}}{dt^{m}}\frac{\partial
L}{\partial x^{\left(  k\right)  }}\right)  +x^{\left(  1\right)  }\sum
_{k=0}^{n}\left(  -\right)  ^{k}\frac{d^{k}}{dt^{k}}\frac{\partial L}{\partial
x^{\left(  k\right)  }}%
\]
Using the equation of motion,
\[
\sum_{k=0}^{n}\left(  -\right)  ^{k}\frac{d^{k}}{dt^{k}}\frac{\partial
L}{\partial x^{\left(  k\right)  }}=0
\]
the final sum vanishes and we have the conserved energy
\[
E=\sum_{k=0}^{n}\sum_{m=0}^{k-1}\left(  -\right)  ^{m}\left(  x^{\left(
k-m\right)  }\frac{d^{m}}{dt^{m}}\frac{\partial L}{\partial x^{\left(
k\right)  }}\right)  -L
\]
The $n=3$ case of this result is given in \cite{Bouda} and elsewhere.

Directly from the generalized equation we see immediately that if a coordinate
$x$ is cyclic, (i.e. $\partial L/\partial x=0$), we still get a conserved
momentum,
\[
p=\sum_{m=0}^{n-1}\left(  -\right)  ^{m+1}\frac{d^{m}}{dt^{m}}\frac{\partial
L}{\partial x^{\left(  m+1\right)  }}%
\]
This follows from
\begin{align}
0 &  =\sum_{k=0}^{n}\left(  -\right)  ^{k}\frac{d^{k}}{dt^{k}}\frac{\partial
L}{\partial x^{\left(  k\right)  }}\\
&  =\sum_{k=1}^{n}\left(  -\right)  ^{k}\frac{d^{k}}{dt^{k}}\frac{\partial
L}{\partial x^{\left(  k\right)  }}\\
&  =\frac{d}{dt}\sum_{m=0}^{n-1}\left(  -\right)  ^{m+1}\frac{d^{m}}{dt^{m}%
}\frac{\partial L}{\partial x^{\left(  m+1\right)  }}\\
&  =\frac{dp}{dt}%
\end{align}
With higher order Lagrangians, there are additional possibilities. Suppose the
lowest $m<n$ partials of $L$ vanish:
\[
\frac{\partial L}{\partial x^{\left(  k\right)  }}=0,\hspace{0.33in}%
k=0,1,\ldots,m-1
\]
Then the sum in the field equation starts at $m,$ and extracting $m$ time
derivatives
\[
0=\frac{d^{m}}{dt^{m}}\left(  \sum_{k=m}^{n}\left(  -\right)  ^{k}%
\frac{d^{k-m}}{dt^{k-m}}\frac{\partial L}{\partial x^{\left(  k\right)  }%
}\right)
\]
so that the momentum
\[
p_{m}^{m}=\frac{d^{m-1}}{dt^{m-1}}\left(  \sum_{k=m}^{n}\left(  -\right)
^{k}\frac{d^{k-m}}{dt^{k-m}}\frac{\partial L}{\partial x^{\left(  k\right)  }%
}\right)
\]
is conserved. Integrating $m-1$ more times,
\[
\sum_{k=0}^{m-1}\frac{1}{k!}p_{m}^{k}t^{k}=\sum_{k=m}^{n}\left(  -\right)
^{k}\frac{d^{k-m}}{dt^{k-m}}\frac{\partial L}{\partial x^{\left(  k\right)  }}%
\]
where we now have $m$ constants, $p_{m}^{k}.$

Higher order systems also permit a Hamiltonian formulation. Let $n=2m-1$ be
any odd integer. We divide the time derivatives of $x$ into even and odd
order, and replace the odd time derivatives with conjugate momenta. For even
$n,$ Hamilton's equations will be supplemented by one additional
Euler-Lagrange equation. Thus, let
\begin{align*}
y_{k}  &  =\frac{d^{2k}x}{dt^{2k}}=x^{\left(  2k\right)  }\\
p_{k}  &  =\frac{\partial L}{\partial x^{\left(  2k+1\right)  }}%
\end{align*}
for $k=0,1,\ldots,m-1.$ The Legendre transformation is employed in the usual
way to express the Hamiltonian in terms of $y_{k}$ and the $p_{k},$%
\[
H\left(  y_{k},p_{k}\right)  =\sum_{k=1}^{n}p_{k}\frac{dy_{k}}{dt}-L
\]
As usual, $H$ is independent of the odd accelerations, $x^{\left(
2k+1\right)  },$ since
\[
\frac{\partial H}{\partial x^{\left(  2k+1\right)  }}=\frac{\partial
H}{\partial\dot{y}}=p_{k}-\frac{\partial L}{\partial x^{\left(  2k+1\right)
}}=0
\]
The variation of the Lagrangian with respect to $y_{k}$ and $p_{k}$ is
straightforward:
\begin{align*}
0  &  =\delta S\\
&  =\int\sum_{k=1}^{n}p_{k}\delta\dot{y}+\sum_{k=1}^{n}\dot{y}\delta
p_{k}-\sum_{k=1}^{n}\frac{\partial H}{\partial y_{k}}\delta y_{k}-\sum
_{k=1}^{n}\frac{\partial H}{\partial p_{k}}\delta p_{k}%
\end{align*}
Integrating by parts we find $2m=n+1$ first order equations:
\begin{align*}
\frac{dy_{k}}{dt}  &  =\frac{\partial H}{\partial p_{k}}\\
\frac{dp_{k}}{dt}  &  =-\frac{\partial H}{\partial y_{k}}%
\end{align*}
These are the generalized Hamilton's equations.

Naturally, higher order equations require more initial data than we usually
have to specify to determine the motion of a classical system, so their
occurrence is somewhat rare. But ultimately, any restriction to second order
equations in classical physics is phenomenological, depending principally on
the success of second order models for fitting measurements. If we take
quantum physics into account, higher order equations in field theory may
introduce ghosts or other undesirable features.

Nonetheless, there are situations where higher order equations are justified.
We briefly discuss one of these below, Bohmian quantum mechanics. The KdV
equation provides an example of an integrable system, having an infinity of
independent constants of motion. It is presented in an Appendix, as is the
approach to chaos.

\subsection{Bohmian quantum mechanics}

A central theme of the Bohmian approach to quantum mechanics is to give it a
form which may be interpreted classically (\cite{Bohm52},\cite{BohmVigier}%
,\cite{Messiah}). The first step is to replace the complex wave function by
pair of real valued functions. This is accomplished as follows. Let
\[
\psi=Ae^{\frac{i}{\hbar}S}%
\]
where $A$ and $S$ are real valued functions. Substituting into the
Schr\"{o}dinger equation,
\[
-\frac{\hbar^{2}}{2m}\nabla^{2}\psi+V\psi=i\hbar\frac{\partial\psi}{\partial
t}%
\]
and separating the real and imaginary parts give two equations:
\begin{align}
\frac{1}{2m}\vec{\nabla}S\cdot\vec{\nabla}S+V+\frac{\partial S}{\partial t}
&  =\frac{\hbar^{2}}{2m}\frac{1}{A}\nabla^{2}A\label{Real part}\\
\frac{\partial A}{\partial t}+\frac{1}{m}\vec{\nabla}S\cdot\vec{\nabla}%
A+\frac{1}{2m}A\nabla^{2}S  &  =0 \label{Imaginary part}%
\end{align}
We refer to this system as the Bohm equations. The first equation is the
Hamilton-Jacobi equation with an additional term. This equation is frequently
used to show how the classical limit emerges from quantum mechanics when
$\hbar\rightarrow0,$ but here we want to exactly replicate the content of the
quantum theory while maintaining a classical viewpoint. Multiplying the second
equation, eq.(\ref{Imaginary part}), by $A,$ it may be rewritten as
\begin{equation}
\frac{\partial A^{2}}{\partial t}+\frac{1}{m}\vec{\nabla}\cdot\left(
A^{2}\vec{\nabla}S\right)  =0 \label{Continuity}%
\end{equation}
Defining
\[
\rho=A^{2},\text{ \ }\mathbf{v}=\frac{1}{m}\vec{\nabla}S
\]
the equation becomes the continuity equation for a current $\mathbf{J}%
=\rho\mathbf{v,}$%
\[
\frac{\partial\rho}{\partial t}+\vec{\nabla}\cdot\mathbf{J}=0
\]
This is the usual conserved probability current of quantum mechanics, cast in
classical guise.

Alternatively, we may view eq.(\ref{Real part}) as a wave equation for $A:$%
\[
-\frac{\hbar^{2}}{2m}\nabla^{2}A+\left(  \frac{1}{2m}\vec{\nabla}S\cdot
\vec{\nabla}S+V+\frac{\partial S}{\partial t}\right)  A=0
\]
This is just a diffusion equation with a horrible potential.

In the one dimensional, stationary case we can reduce the Bohm equations to a
single, higher order differential equation. In $1$-dim, with the stationary
conditions
\[
\frac{\partial S}{\partial t}=-E,\text{ \ \ }\frac{\partial A}{\partial t}=0
\]
the continuity equation, eq.(\ref{Continuity}) is simply
\[
\frac{1}{m}\left(  A^{2}S^{\prime}\right)  ^{\prime}=0
\]
which integrates immediately to give
\[
A^{2}=\frac{a}{S^{\prime}}%
\]
Now substituting this result into eq.(\ref{Real part}) together with $\partial
S/\partial t=-E,$ results in a third-order, nonlinear equation
\begin{equation}
\frac{\hbar^{2}}{2m}\left[  \frac{S^{\prime\prime\prime}}{2S^{\prime}}%
-\frac{3}{4}\left(  \frac{S^{\prime\prime}}{S^{\prime}}\right)  ^{2}\right]
+\frac{1}{2m}\left(  S^{\prime}\right)  ^{2}+V-E=0\label{Bohm equation}%
\end{equation}
The interesting point is that this equation is rigorously equivalent to the
$1$-dim stationary state Schr\"{o}dinger equation. The downside is that we
have handled only the $1$-dim, stationary case.

This equation lacks the simplicity of the Schr\"{o}dinger equation. For
example, suppose we solve
\[
-\frac{\hbar^{2}}{2m}\left[  \frac{3}{4}\left(  \frac{S^{\prime\prime}%
}{S^{\prime}}\right)  ^{2}-\frac{S^{\prime\prime\prime}}{2S^{\prime}}\right]
+\frac{1}{2m}\left(  S^{\prime}\right)  ^{2}+V-E=0
\]
to find a solution $S\left(  x,E\right)  .$ Unfortunately, the nonlinearity
means that we cannot take a superposition of stationary states to get a
general time-dependent solution.

Consider the time-dependent case further. We can at least find a time
dependent solution to the (linear!) second equation:
\[
\frac{\partial A}{\partial t}+\frac{1}{m}S^{\prime}A^{\prime}+\frac{1}%
{2m}AS^{\prime\prime}=0
\]
Direct integration shows that
\[
A\left(  x,t\right)  =\sqrt{\frac{2\pi}{S^{\prime}}}G\left(  t-m\int\frac
{dx}{S^{\prime}}\right)
\]
is a solution for any function $G.$ The amplitude therefore propagates with
fixed spatial form. It would be of interest to know if this type of solution
for $A$ generalizes to higher dimensions. Now we may substitute this into the
first equation,
\[
\frac{\partial S}{\partial t}+\frac{1}{2m}\left(  S^{\prime}\right)
^{2}+V=\frac{\hbar^{2}}{2m}\frac{1}{\sqrt{\frac{2\pi}{S^{\prime}}}G\left(
t-m\int\frac{dx}{S^{\prime}}\right)  }\left[  \sqrt{\frac{2\pi}{S^{\prime}}%
}G\left(  t-m\int\frac{dx}{S^{\prime}}\right)  \right]  ^{\prime\prime}%
\]
Perhaps an appropriate separation can solve this equation as well.

\subsubsection{Bohmian Lagrangian}

The $1$-dim, stationary state Bohm equation also follows from the variation of
the usual Schr\"{o}dinger action, reduced by the solution for $A$. Starting
from the Schr\"{o}dinger action,
\[
S=\int\frac{\hbar^{2}}{2m}\left(  \psi^{\ast}\right)  ^{\prime}\psi^{\prime
}+V\psi^{\ast}\psi+\frac{i}{2}\hbar\left(  \frac{\partial\psi^{\ast}}{\partial
t}\psi-\frac{\partial\psi}{\partial t}\psi^{\ast}\right)
\]
we substitute the polar expression $\psi=\frac{1}{\sqrt{S^{\prime}}}%
e^{\frac{i}{\hbar}[S\left(  x\right)  -Et]}$ for the wave function. the action
becomes
\[
S=\int\frac{\hbar^{2}}{2m}\frac{\left(  S^{\prime\prime}\right)  ^{2}%
}{4\left(  S^{\prime}\right)  ^{3}}+\frac{1}{2m}S^{\prime}+\frac
{V-E}{S^{\prime}}%
\]
Varying, the equation of motion is found to be
\[
0=\frac{d}{dx}\left[  -\frac{\hbar^{2}}{2m}\frac{2S^{\prime\prime\prime}%
}{4\left(  S^{\prime}\right)  ^{3}}+\frac{\hbar^{2}}{2m}\frac{3S^{\prime
\prime}S^{\prime\prime}}{2\left(  S^{\prime}\right)  ^{4}}-\frac{\hbar^{2}%
}{2m}\frac{3\left(  S^{\prime\prime}\right)  ^{2}}{4\left(  S^{\prime}\right)
^{4}}+\frac{1}{2m}-\frac{V-E}{\left(  S^{\prime}\right)  ^{2}}\right]
\]
The term in brackets is a constant. Choosing the constant to be $1/m$
correctly reproduces the stationary Bohm equation:
\[
-\frac{\hbar^{2}}{4m}\frac{S^{\prime\prime\prime}}{S^{\prime}}+\frac
{3\hbar^{2}}{8m}\left(  \frac{S^{\prime\prime}}{S^{\prime}}\right)  ^{2}%
-\frac{\left(  S^{\prime}\right)  ^{2}}{2m}-\left(  V-E\right)  =0
\]
We could have avoided the need to pick this integration constant by taking the
action to be%

\[
\tilde{S}=\int\frac{\hbar^{2}}{2m}\frac{\left(  S^{\prime\prime}\right)  ^{2}%
}{4\left(  S^{\prime}\right)  ^{3}}-\frac{1}{2m}S^{\prime}+\frac{V-E}{
S^{\prime}}%
\]
This form differs from the previous one only by the integral of a total derivative.

Writing the equation of motion in the form,
\begin{equation}
0=-\frac{\hbar^{2}}{4m}S^{\prime}S^{\prime\prime\prime}+\frac{3\hbar^{2}}%
{8m}\left(  S^{\prime\prime}\right)  ^{2}-\frac{1}{2m}\left(  S^{\prime
}\right)  ^{4}-\left(  V-E\right)  \left(  S^{\prime}\right)  ^{2}
\label{Floyd}%
\end{equation}
it has been observed (\cite{Floyda},\cite{Floydb}) that there are solutions
with $S^{\prime}=0.$ It is not difficult to see that these should not be
considered to be the physical solutions. First, we arrived at eq.(\ref{Floyd})
by the substitution
\[
A=1/\sqrt{S^{\prime}}%
\]
which is singular for constant $S.$ Since wave functions with divergent
amplitude are not considered physical, such points require a closer
examination at the very least. Second, we can see that these are clearly not
the physical solutions because they are independent of the potential $V.$

To eliminate the spurious solutions, while indicating a method of solution, we
make a simple change of variable. Notice that the nonlinear terms in the Bohm
equation, eq.(\ref{Bohm equation}), may be rewritten as
\[
-2\sqrt{S^{\prime}}\left(  \frac{1}{\sqrt{S^{\prime}}}\right)  ^{\prime\prime
}=\frac{S^{\prime\prime\prime}}{S^{\prime}}-\frac{3}{2}\left(  \frac
{S^{\prime\prime}}{S^{\prime}}\right)  ^{2}%
\]
Substituting, the equation of motion becomes
\[
-\frac{\hbar^{2}}{2m}\left(  \frac{1}{\sqrt{S^{\prime}}}\right)
^{\prime\prime}+\frac{\left(  S^{\prime}\right)  ^{3/2}}{2m}+\frac{V-E}%
{\sqrt{S^{\prime}}}=0
\]
Now let $r=1/\sqrt{S^{\prime}}.$ Then
\[
-\frac{\hbar^{2}}{2m}r^{\prime\prime}+\frac{1}{2mr^{3}}+\left(  V-E\right)
r=0
\]
Rearranging,
\[
mr^{\prime\prime}-\frac{m^{2}/\hbar^{2}}{mr^{3}}=\frac{2m^{2}}{\hbar^{2}%
}\left(  V-E\right)  r
\]
the form suggest that we think of the independent variable $x$ as time.
Replacing $x\rightarrow t,$ we interpret $r\left(  t\right)  $ as a radial
coordinate and $V$ as a time-dependent potential $V\left(  t\right)  .$ Then
we have
\[
m\ddot{r}-\frac{m^{2}/\hbar^{2}}{mr^{3}}=\frac{2m^{2}}{\hbar^{2}}[V\left(
t\right)  -E]r
\]
and we recognize the isotropic $2$-dim harmonic oscillator with angular
momentum $m/\hbar$ with a time-dependent spring strength,
\[
k\left(  t\right)  =-\frac{2m^{2}}{\hbar^{2}}[V\left(  t\right)  -E]
\]
This clearly has bound state solutions for suitable energies and potentials.
Notice that solutions with $S^{\prime}=0$ correspond to infinite radial
coordinate, so bound state solutions automatically avoid this spurious case.

It is suggestive that the introduction of a time-dependent spring constant
into the isotropic oscillator can lead to parametric resonance
\cite{LandauLifshitz}. It would be of interest to find the relationship of
such parametric resonances to the eigenmodes of the corresponding quantum problem.

\section{Gauging Newton's Law}

One surprising new result in classical mechanics is that both the Lagrangian
and Hamiltonian formulations of Newton's laws may be derived as gauge theories
of Newton's second law (\cite{WheelerGaugingNewton},\cite{AndersonWheeQuant}).
To see how this comes about, and to understand the symmetries involved, we
digress a moment to consider the essential elements of a physical theory. In
particular, we want to distinguish two features: dynamical laws and
measurement theory.

The distinction between these is easy to see. For example, in quantum
mechanics the dynamical law is the Schr\"{o}dinger equation
\[
\hat{H}\psi-i\hbar\frac{\partial\psi}{\partial t}%
\]
which governs the time evolution of the wave function, $\psi.$ The measurement
theory is what establishes the correspondence between calculations and
measurable numbers. One of the chief elements of quantum measurement theory is
therefore the Hermitian inner product on Hilbert space:
\[
\left\langle \psi|\psi\right\rangle =\int_{V}\psi^{\ast}\psi d^{3}x.
\]
As a second example, consider Newtonian mechanics. The dynamical variable for
a particle is the position vector, $\mathbf{x},$ and its motion is governed by
the second law:
\[
\mathbf{F}=m\frac{d^{2}\mathbf{x}}{dt^{2}}%
\]
while the inner product allows us to extract measurable magnitudes
\[
\left\langle \mathbf{u},\mathbf{v}\right\rangle =\mathbf{u}\cdot\mathbf{v}%
\]

We are intested in the differences in the \textit{symmetries} of dynamical
laws and measurement. Generally, the differential or other equation governing
dynamical evolution is invariant under some global symmetry. In contrast to
this, a metric, an invariant product or some other real-valued mapping to
measurable quantities is often invariant under a group of diffeomorphisms.
Whatever the symmetries, it often occurs that the symmetry of dynamical
evolution and the symmetry of the measurement theory are different. Which is
the symmetry of the system?

For Newtonian measurements, the inner products allow local transformations and
therefore have the larger symmetry. It makes sense to try to extend the
symmetry of the dynamical law to agree with that of the measurement theory.
Fortunately, there are standard techniques for accomplishing this extension --
the methods of gauge theory.

\textit{Gauging} takes a global symmetry, that is, a symmetry that is
independent of position and time, and extends it to a local symmetry, i.e.,
one that may be different at different positions. We systematically extend to
a local symmetry by introducing a connection: a one-form field valued in the
Lie algebra of the symmetry we wish to gauge. Added to the usual partial
derivative, the connection subtracts back out the extra terms arising from
differentiating the local symmetry. The most familiar example is general
relativity, in which the Christoffel connection, $\Gamma_{\mu\nu}^{\alpha}$
added to partial derivative makes the derivative covariant with respect to
general coordinate transformations. The $U(1)$ gauge theory of
electromagnetism is also familiar. In this case, the vector potential provides
the connection.

What are the symmetries of the Newtonian dynamical and measurement theories?
There is more than one answer. The dynamical law is invariant under the
Galilean group, $G,$ consisting of rotations, translations, Galilean boosts
and time translations. It is possible to extend the rotations to general
linear transformations and still leave the second law invariant. For the
measurement theory, the Euclidean line element is invariant under the set of
3-dimensional rotations and translations, $ISO(3)$. This is called the
\textit{Euclidean group}. If we regard Euclidean 3-space as a manifold instead
of a vector space, these transformations may be local. Furthermore,
recognizing that we only actually measure dimensionless ratios (for example,
the ratio of the height of a tree to the length of a meter stick), we can
require invariance of ratios$\ $of line elements. This gives the conformal
group, $SO(4,1).$

We will consider two of the possible gaugings of Newton's second law --
Euclidean and conformal.

\subsection{Euclidean gauge theory of Newton's second law}

The Euclidean gauging of Newton's second law leads to Lagrangian mechanics.
This is not a particularly surprising result. However, while it is well-known
that Lagrangian mechanics provides a form of Newton's second law valid for
``generalized coordinates,'' the construction by gauging has advantages: $(1)$
it arrives at the Lagrangian formulation in a systematic way, $(2)$ gauging
displays explicitly the meaning of generalized coordinates, and $(3)$ it
illustrates the general techniques used for the gauging of conformal symmetry
below. As we shall see, the results of conformal gauging in the next
subsection are unexpected.

Proceeding, first recall that the transformations of the Euclidean group
$ISO(3)$ include three rotations and three translations, and the Lie algeba
has corresponding generators. Gauging therefore gives us two sets of $1$-form
gauge fields:

\begin{enumerate}
\item Three translational gauge fields, comprising the dreibein,
$\mathbf{e}^{i}.$

\item Three rotational gauge fields, the $SO(3)$ spin connection,
$\mathbf{\omega}_{j}^{i},$ antisymmetric under the interchange of indices.
\end{enumerate}

The gauging proceeds just as when we gauge the Poincar\'{e} group to develop
Riemannian geometry (\cite{Utiyama},\cite{Kibble}). In Poincar\'{e} gauging,
the vierbein, $\mathbf{e}^{a},$ is identified with an orthonormal frame field
on a $4$-dim Riemannian manifold and the spin connection $\mathbf{\omega}%
_{b}^{a}$ permits the use of local Lorentz transformations. For the present
Euclidean case, the dreibein, $\mathbf{e}^{i},$ is identified with an
orthonormal basis of a $3$-dim manifold and the $SO(3)$ spin connection,
$\mathbf{\omega}_{j}^{i},$ gives local rotational symmetry. The pair $\left(
\mathbf{e}^{i},\mathbf{\omega}_{j}^{i}\right)  $ is equivalent to the metric
and general coordinate connection, $\left(  g_{mn},\Gamma_{rs}^{m}\right)  .$

The connection forms must satisfy the Lie algebra relations of the symmetry
group, as encoded in the Maurer-Cartan structure equations:
\begin{align*}
\mathbf{d\omega}_{j}^{i}  &  =\mathbf{\omega}_{j}^{k}\mathbf{\omega}_{k}^{i}\\
\mathbf{de}^{i}  &  =\mathbf{e}^{j}\mathbf{\omega}_{j}^{i}%
\end{align*}
The solution of these is simple since we do not include curvature. The first
equation is solved by the pure gauge form of the connection,
\[
\mathbf{\omega}_{k}^{i}=\left(  \mathbf{d}\Lambda_{j}^{i}\right)  \left[
\Lambda^{-1}\right]  _{k}^{j}%
\]
where $\Lambda_{j}^{i}\left(  \mathbf{x}\right)  $ is a local rotation matrix.
This means that there exists a choice of frames (say, $\Lambda_{j}^{i}=$
constant) in which the spin connection is zero. Choosing this frame, the
equation for the dreibein is satisfied by setting
\[
\mathbf{e}^{i}=\mathbf{d}x^{i}%
\]
From this we see that the equations describe Euclidean $3$-space. Using the
spin connection we can define a derivative operator which is covariant with
respect to local rotations. If we cast the same equations in terms of a
coordinate basis using the metric and Christoffel connection, $\left(
g_{mn},\Gamma_{rs}^{m}\right)  ,$ the derivative is covariant with respect to
general coordinate changes, or diffeomorphisms.

We may find the new dynamical law using a variational principle. Using the
coordinate metric,
\[
g_{mn}=e_{m}^{\ \ i}e_{n}^{\ \ j}\delta_{ij}%
\]
we choose the squared norm of the velocity vector, plus a function of the
coordinates to provide a source for the motion:
\[
S=\int[g_{mn}v^{m}v^{n}+\phi\left(  x^{m}\right)  ]dt
\]
Because we have local symmetry, we can write the same thing in any
coordinates. Notice that there is always some arbitrariness in the gauging
procedure at this point. There are two properties we demand of this
variational principle. First, it must be invariant under the local symmetry
group. Second, we require the restriction of the new dynamical law to the
original symmetry to reproduce the original law. The action $S$ above
satisfies these requirements.

Varying $S,$ we find the new form of Newton's law,
\[
g_{mn}\frac{Dv^{n}}{dt}=\frac{\partial\phi}{\partial x^{m}}%
\]
where the covariant derivative of $v^{m}$ transforms as a vector under local
rotations. When $\phi=0,$ this is the geodesic equation. Since the space is
Euclidean, the geodesics are straight lines. The class of straight lines
\[
x^{\alpha}=x_{0}^{\alpha}+v_{0}^{\alpha}t
\]
is equivalent to the class of Newtonian inertial reference frames.

Writing $V=-\frac{m}{2}\varphi$ for a potential $V,$ we see that forces
produce deviations from geodesic motion. This is the Lagrangian formulation of
mechanics. Note that we get the same equation of motion if we substitute the
Lagrangian in the form
\[
L=g_{mn}v^{m}v^{n}+\phi\left(  x^{m}\right)
\]
into the usual Euler-Lagrange equation. The general coordinate invariance
(``use of generalized coordinates'') is, of course, one of the main reasons
for the use of Lagrangian methods. The present approach, while principally
intended to pave the way for the conformal gauging below, does have the
advantage of systematically showing that the class of generalized coordinates
is just the diffeomorphism group. In the usual formulation, this conclusion
follows from the coordinate invariance of the action.

\subsection{Conformal gauge theory of Newton's second law}

We now repeat the gauging process, but this time use the full conformal
symmetry. The conformal group (for compactified $3$-dim Euclidean space)
contains ten transformations:

\begin{enumerate}
\item 3 rotations

\item 3 translations

\item 1 dilatation

\item 3 special conformal transformations
\end{enumerate}

The first two sets of transformations reproduce the Euclidean group.
Dilatations just rescale all lengths by a factor, while special conformal
transformations are translations in inverse coordinates.

These global transformations preserve the Euclidean line element up to an
overall multiple. As it stands, Newton's second law is not invariant under
even the \textit{global} form of these transformations -- the special
conformal transformations do not leave the law unchanged because they do not
act linearly on Euclidean $3$-space. This is easy to fix: we introduce a very
limited covariant derivative with a connection specific to global special
conformal transformations. It is unusual to require a connection in a
dynamical law before gauging, but nothing forbids it and it gives us an
equation with the symmetry we wish to gauge.

Now consider Newton's law, modified just enough to let us perform all $10$
global conformal transformations. Make those $10$ global transformations
local. There is more than one way to do this, but so far only one appears to
be interesting -- the \textit{biconformal} gauging described below.

There will now be ten gauge fields:

\begin{enumerate}
\item The dreibein, $\mathbf{e}^{i}$

\item The (antisymmetric) $SO(3)$ spin connection, $\mathbf{\omega}_{j}^{i} $

\item The Weyl vector, $\mathbf{W}.$

\item The co-dreibein, $\mathbf{f}_{i}$, from special conformal transformations
\end{enumerate}

The local rotations, gauged by the spin connection, are as expected and we add
local dilatations gauged by the Weyl vector. These allow general coordinate
invariance and scale invariance. Employing the biconformal technique, we
interpret $\left(  \mathbf{e}^{i},\mathbf{f}_{i}\right)  $ as an orthonormal
frame field of a \textit{six} dimensional space.

These gauge fields must satisfy the Maurer-Cartan structure equations:
\begin{align*}
\mathbf{d\omega}_{j}^{i}  &  =\mathbf{\omega}_{j}^{k}\mathbf{\omega}_{k}%
^{i}+\mathbf{e}^{i}\mathbf{f}_{j}-\mathbf{e}_{j}\mathbf{f}^{i}\\
\mathbf{de}^{i}  &  =\mathbf{e}^{j}\mathbf{\omega}_{j}^{i}+\mathbf{We}^{i}\\
\mathbf{df}_{i}  &  =\mathbf{\omega}_{i}^{j}\mathbf{f}_{j}+\mathbf{f}%
_{i}\mathbf{W}\\
\mathbf{dW}  &  =2\mathbf{e}^{i}\mathbf{f}_{i}%
\end{align*}
This is just the conformal Lie algebra in a dual basis. Once again the
equations are easily solved. The solution reveals a symplectic form,
\begin{align*}
\theta &  =\mathbf{e}^{k}\mathbf{f}_{k}\\
\mathbf{d}\theta &  =0
\end{align*}
The six dimensional space therefore has a similar structure to a one particle
phase space. The units of the coordinates of this $6$-dim space are not all
the same. Three are correct for postition $(x^{i},length)$ while the remaining
three are geometric units for momentum $(y_{i},1/length).$ Note that the
conversion of units of momentum to units of inverse length may be accomplished
using any conventional dimensional standards, e.g., meters, seconds and
kilograms. It also follows from the solution that the Weyl vector is given by
\[
\mathbf{W}=-y_{i}\mathbf{d}x^{i}%
\]

To find the new dynamical law we again write an action. Since the geometry is
like phase space, the paths will not be anything like geodesics, so path
length will not work. Instead, we have a new feature -- the Weyl vector --
that comes from the dilatations. We will base our dynamical law on the
geometric interpretation of this vector field. We digress briefly to explore
its properties.

It follows from the nature of conformal geometry that the integral of the Weyl
vector along any path gives the relative physical size change along that
path:
\[
l=l_{0}\exp\int W_{i}v^{i}dt=l_{0}\exp\int\mathbf{W}%
\]
This means that magnitudes are not preserved -- initially identical rods
transported along different curves might be different sizes when they are
returned together and compared. This possibility is the price we pay for the
freedom to make local scale transformations, just as in a Riemannian geometry
vectors may rotate even under ``parallel'' transport. We will return to this
point below.

We take the action to be the integral of the Weyl vector. Then the physical
paths will be paths of extremal size change. Notice that, while the
exponential above is gauge dependent, its variation is not. Indeed, it is
worth noting that the gauge freedom of the Weyl vector agrees exactly with the
freedom to add a total derivative to a Lagrangian.

Once again we add a function to provide a source,
\[
S=\int\left(  \mathbf{W}\cdot\mathbf{v}+\phi\right)  dt
\]
It is interesting that such a function is provided automatically in the
relativistic version of this gauging, as the time component of the Weyl vector.

Now vary the action. There are six first-order equations:
\begin{align*}
\frac{dx^{i}}{dt}  &  =\frac{\partial\phi}{\partial y_{i}}\\
\frac{dy_{i}}{dt}  &  =-\frac{\partial\phi}{\partial x^{i}}%
\end{align*}
If we identify $\phi$ with the Hamiltonian, these are Hamilton's equations.
Therefore, the gauge theory of Newton's second law with respect to the
conformal grouip is Hamiltonian mechanics.

There are a couple of points to be clarified. First, the multiparticle case
works even though a single Weyl vector must account for the Hamiltonian and
momentum of each particle as long as we assume that two particles never occupy
exactly the same space. This is consistent with the usual requirements of
Newtonian mechanics, by which matter is impenetrable.

Second, the extremal value of the integral of the Weyl vector is zero. Thus,
no measurable size change occurs for \textit{classical} motion, even though
the Weyl vector does not have vanishing curl. The classical paths are
precisely the ones along which no physical dilatation is ever measured.

It is also interesting to note that there is a $6$-dim metric, of an
unexpected form that is consistent with collisions. It follows from the
solution to the structure equations that the line element is of the form
\[
ds^{2}=dx^{i}dx^{i}+dx^{i}dy_{i}%
\]
Therefore, if we assume that the distance $ds$ between two particles must
vanish (or nearly so) in order for two particles to collide, we see we must
have $dx^{i}=0,$ regardless of their separation in momentum, $dy_{i}$. This
would not be the case if we had simply imposed a Euclidean metric on the space.

These results provide a satisfying unification of classical mechanics. In
addition, the relativistic version of biconformal gauging also turns out to be
interesting. We have shown (\cite{WhJMP},\cite{WW}) that the method provides
the best way to understand conformally invariant gravity. The results are
consistent with general relativity, and improve previous conformal gravity
theories. The fact that we can satisfactorily gauge classical mechanics -- and
get something new -- gives us a better understanding of, and more confidence
in, the relativistic theory.

There is also a suggestion of something deeper. Notice that quantum mechanics
\textit{requires} both position and momentum variables to make sense, while
biconformal gauging of Newton's second law gives us a space which
automatically has both sets of variables. Is it possible that quantum physics
takes on a particularly simple form in biconformal space? Anderson and Wheeler
claim it does, deriving a path integral formulation of quantum mechanics
directly from a biconformal measurement theory \cite{AndersonWheeQuant}. It
becomes possible to claim that the physical manifold is really a six- (or,
relativistically, an eight-) dimensional place, in which quantum mechanics is
a natural description of phenomena.

This interpretation of biconformal space works correctly. In particular, when
we use the covering group of the conformal group, the Weyl vector is
necessarily complex. The presence of an $``i"$ in the Weyl vector makes an
initially real, probabilistic evolution law into a unitary evolution. In
addition, the requirement of the scale-invariant theory for taking ratios of
lengths to produce a meaningful measurement leads directly to the use of the
product of probability amplitudes in computing physically measurable
probabilities. Naturally, the proportionality between the inverse-length
$y_{i}$-coordinates and momenta is taken to be
\[
\hbar y_{i}=p_{i}%
\]
Note that this factor drops out of Hamilton's equations, making Planck's
constant classically unmeasurable.

Thus, we have another way to think of quantum phenomena in a classical
context. In this formulation, however, we have the added advantage of a direct
connection to general relativity. It becomes possible to ask questions about
quantum measurement of curved spacetimes in a classical context.

\section{Spin, statistics, and pseudomechanics in classical physics}

\subsection{Spin}

Now that we have a gauge theory of mechanics, we can ask further about the
representation of the gauge symmetry. A representation of a group is the
vector space on which the group acts. The largest class of objects on which
our symmetry acts will be the class determining the \textit{covering group}.
This achieves the fullest realization of our symmetry. For example, while the
Euclidean group $ISO\left(  3\right)  $ leads us to the usual formulation of
Lagrangian mechanics, we can ask if we might not achieve something new by
gauging the covering group, $ISpin(3)\cong ISU\left(  2\right)  .$ This
extension, which places spinors in the context of classical physics, depends
only on symmetry, and therefore is completely independent of quantization.

There are numerous advantages to the spinorial extension of classical physics.
After Cartan's discovery of spinors as linear representations of orthogonal
groups in 1913 (\cite{Cartan13},\cite{Cartan66}) and Dirac's use of spinors in
the Dirac equation (\cite{Dirac28},\cite{Dirac30}), the use of spinors for
other areas of relativistic physics was pioneered by Penrose (\cite{Penrose60}%
,\cite{PenroseRindler}). Penrose developed spinor notation for general
relativity that is one of the most powerful tools of the field. For example,
the use of spinors greatly simplifies Petrov's classification of spacetimes
(compare Petrov \cite{Petrov} and Penrose \cite{Penrose60},\cite{Wald}), and
tremendously shortens the proof of the positive mass theorem (compare Schoen
and Yau (\cite{SchoenYau},\cite{SchoenYau81},\cite{SchoenYau82}) and Witten
\cite{WittenPosMass}). Penrose also introduced the idea and techniques of
\textit{twistor spaces}. While Dirac spinors are representations of the
Lorentz symmetry of Minkowski space, twistors are the spinors associated with
larger conformal symmetry of compactified Minkowski space. Their overlap with
string theory as twistor strings is an extremely active area of current
research in quantum field theory (see \cite{Witten} and references thereto).
In nonrelativistic classical physics, the use of Clifford algebras (which,
though they do not provide a spinor representation in themselves, underlie the
definition of the spin groups) has been advocated by Hestenes in the
\textquotedblleft geometric algebra\textquotedblright\ program \cite{Hestenes}.

It is straightforward to include spinors in a classical theory. We provide a
simple example. For the rotation subgroup of the Euclidean group, we can let
the group act on complex $2$-vectors, $\chi^{a}$, $a=1,2.$ The resulting form
of the group is $SU(2).$ In this representation, an ordinary $3$-vector such
as the position vector $x^{i}$ is written as a traceless Hermitian matrix,
\begin{align*}
X  &  =x^{i}\sigma_{i}\\
\left[  X\right]  ^{ab}  &  =x^{i}\left[  \sigma_{i}\right]  ^{ab}%
\end{align*}
where $\sigma_{i}$ are the Pauli matrices. It is easy to write the usual
Lagrangian in terms of $X:$%
\[
L=\frac{m}{4}tr\left(  \dot{X}\dot{X}\right)  -V\left(  X\right)
\]
where $V$ is any scalar-valued function of $X.$ However, we now have the
additional complex $2$-vectors, $\chi^{a},$ available. Consider a Dirac-type
kinetic term
\[
\lambda\chi_{a}\left(  i\dot{\chi}^{a}-\mu\chi^{a}\right)
\]
and potential
\[
V\left(  \chi^{a}\right)  =\lambda\bar{\chi}^{a}B^{i}\sigma_{iab}\chi
^{b}+\ldots
\]
Notice there is no necessity to introduce fermions and the concomitant
anticommutation relations -- we regard these spinors as commuting variables. A
simple action therefore takes the form
\[
S=\int dt\left[  \frac{m}{4}tr\left(  \dot{X}\dot{X}\right)  +\bar{\chi}%
_{a}\left(  i\dot{\chi}^{a}-\mu\chi^{a}\right)  -V\left(  X\right)
-\lambda\bar{\chi}^{a}B^{i}\sigma_{iab}\chi^{b}\right]
\]
The equations of motion are then
\begin{align*}
m\ddot{x}^{i}  &  =-\sigma^{iab}\frac{\partial V}{\partial X^{ab}}\\
\dot{\chi}^{a}  &  =-i\mu\chi^{a}-i\lambda B^{i}\sigma_{iab}\chi^{b}%
\end{align*}
together with the complex conjugate of the second. The first reproduces the
usual equation of motion for the position vector. Assuming a constant vector
$B^{i},$ we can easily solve the second. Setting $\chi=\psi e^{-i\mu t},$
$\psi$ must satisfy
\[
\dot{\psi}=-i\lambda B^{i}\sigma_{i}^{\ ab}\psi_{b}%
\]
This describes steady rotation of the spinor,
\[
\psi=e^{-i\lambda B}\psi_{0}%
\]
The important thing to note here is that, while the spinors $\psi$ rotate with
a single factor of $e^{i\mathbf{w}\cdot\sigma},$ a vector such as $X$ rotates
as a matrix and therefore requires two factors of the rotation
\[
X^{\prime}=e^{-i\mathbf{w}\cdot\sigma}Xe^{i\mathbf{w}\cdot\sigma}%
\]
This illustrates the $2:1$ ratio of rotation angle characteristic of spin
$1/2.$ The new degrees of freedom therefore describe classical spin and we see
that spin is best thought of as a result of the symmetries of classical
physics, rather than as a necessarily quantum phenomenon. Similar results
using the covering group of the Lorentz group introduce Dirac spinors
naturally into relativity theory. Indeed, as noted above, $2$-component spinor
notation is a powerful tool in general relativity, where it makes such results
as the Petrov classification or the positivity of mass transparent.

\subsection{Statistics and pseudomechanics}

The use of spinors brings immediately to mind the exclusion principle and the
spin-statistics theorem. We stressed that spin and statistics are independent.
Moreover, spin, as described above, follows from the use of the covering group
of any given orthogonal group and is therefore classical. For statistics, on
the other hand, the situation is not so simple. In quantum mechanics, the
difference between Bose-Einstein and Fermi-Dirac statistics is a consequence
of the combination of anticommuting variables with the use of discrete states.
In classical physics we do not have discrete states. However, nothing prevents
us from introducing anticommuting variables. In its newest form, the resulting
area of study is called \textit{pesudomechanics}.

The use of anticommuting, or Grassmann variables in classical physics actually
has an even longer history than spin. The oldest and most ready example is the
use of the wedge product for differential forms
\[
\mathbf{d}x\wedge\mathbf{d}y=-\mathbf{d}y\wedge\mathbf{d}x
\]
This gives a grading of $\left(  -\right)  ^{p}$ to all $p$-forms. Thus, if
$\omega$ is a $p$-form and $\eta$ a $q$-form,
\begin{align*}
\mathbf{\omega}  &  =\omega_{i_{1}\cdots i_{p}}\mathbf{d}x^{i_{i}}\wedge
\cdots\wedge\mathbf{d}x^{i_{p}}\\
\mathbf{\eta}  &  =\omega_{i_{1}\cdots i_{q}}\mathbf{d}x^{i_{1}}\wedge
\cdots\wedge\mathbf{d}x^{i_{q}}%
\end{align*}
Then their (wedge) product is even or odd depending on whether $pq$ is even or
odd:
\[
\mathbf{\omega}\wedge\mathbf{\eta}=\left(  -\right)  ^{pq}\mathbf{\eta}%
\wedge\mathbf{\omega}%
\]
Nonetheless, $p$-forms rotate as covariant, rank-$p$ tensors under $SO(3)$ (or
$SO\left(  n\right)  $), in violation of the familiar spin-statistics theorem.
Under $SU(2)$ they rotate as covariant, rank-$2p$ tensors, not as spinors.

Another appearance of anticommuting variables in classical mechanics stems
from the insights of supersymmetric field theory. Before supersymmetry,
continuous symmetries in classical systems were characterized by Lie algebras,
with each element of the Lie algebra generating a symmetry transformation. The
Lie algebra is a vector space characterized by a closed commutator product and
the Jacobi identity. Supersymmetries are extensions of the normal Lie
symmetries of physical systems to include symmetry generators (Grassmann
variables) that anticommute. Like the grading of differential forms, all
transformations of the graded Lie algebra are assigned a grading, $0$ or $1,$
that determines whether a commutator or commutator is appropriate, according
to
\[
\left[  T_{p},T_{q}\right]  \equiv T_{p}T_{q}-\left(  -\right)  ^{pq}%
T_{q}T_{p}%
\]
where $p,q\in\left\{  0,1\right\}  .$ Thus, two transformations which both
have grading $1$ have anticommutation relations with one another, while all
other combinations satisfy commutation relations.

Again, there is nothing intrinsically ``quantum'' about such generalized
symmetries, so we can consider classical supersymmetric field theories and
even supersymmetrized classical mechanics. Since anticommuting fields
correspond to fermions in quantum mechanics, we may continue to call variables
fermionic when used classically, even though their statistical properties may
not be Fermi-Dirac. Perhaps more importantly, we arrive at a class of
classical action functionals whose quantization leads directly to Pauli or
Dirac spinor equations.

Casalbuoni pioneered the development of pseudomechanics, showing that it was
possible to formuate an $\hbar\rightarrow0$ limit of a quantum system in such
a way that the spinors remain but their magnitude is no longer quantized
(\cite{Casalbuoni76a}, \cite{Casalbuoni76b}, see also Freund \cite{Freund}).
Conversely, the resulting classical action leads to the Pauli-Schr\"{o}dinger
equation when quantized. Similarly, Berezin and Marinov
\cite{BerezinMarinov77}, and Brink, Deser, Zumino, di Vecchia and Howe
\cite{BrinkDZdVH} introduced four anticommuting variables, $\theta^{\alpha}$
to write the pre-Dirac action. We display these actions below, after giving a
simplified example. Since these approaches moved from quantum fields to
classical equations, they already involved spinor representations. However,
vector versions (having anticommuting variables without spinors) are possible
as well. Our example below is of the latter type. Our development is a slight
modification of that given by Freund \cite{Freund}.

To construct a simple pseudomechanical model, we introduce a superspace
formulation, extending the usual \textquotedblleft bosonic\textquotedblright%
\ $3$-space coordinates $x_{i}$ by three additional anticommuting coordinates,
$\theta^{a},$%
\[
\left\{  \theta^{a},\theta^{b}\right\}  =0
\]
Consider the motion of a particle described by $[x_{i}\left(  t\right)
,\theta^{a}\left(  t\right)  ],$ and the action functional
\[
S=\int dt\left[  \frac{1}{2}m\dot{x}^{i}\dot{x}^{i}+\frac{i}{2}\theta^{a}%
\dot{\theta}^{a}-V\left(  x^{i},\theta^{b}\right)  \right]
\]
Notice that $\theta^{2}=0$ for any anticommuting variable, so the linear
velocity term is the best we can do. For the same reason, the Taylor series in
$\theta^{a}$ of the potential $V\left(  x^{i},\theta^{b}\right)  $
terminates:
\[
V\left(  x^{i},\theta^{b}\right)  =V_{0}\left(  x^{i}\right)  +\psi_{a}\left(
x^{i}\right)  \theta^{a}+\frac{1}{2}\varepsilon_{abc}B^{a}\left(
x^{i}\right)  \theta^{b}\theta^{c}+\frac{1}{3!}\kappa\left(  x^{i}\right)
\varepsilon_{abc}\theta^{a}\theta^{b}\theta^{c}%
\]
Since the coefficients remain functions of $x^{i},$ we have introduced four
new fields into the problem. However, they are not all independent. If we
change coordinates from $\theta^{a}$ to some new anticommuting variables,
setting
\begin{align*}
\theta^{a} &  =\chi^{a}+\xi B_{bc}^{a}\chi^{b}\chi^{c}+C^{a}\varepsilon
_{bcd}\chi^{b}\chi^{c}\chi^{d}\\
B_{bc}^{a} &  =B_{\left[  bc\right]  }^{a}%
\end{align*}
where $\zeta$ is an anticommuting constant, the component functions in
$H\left(  \theta^{b}\right)  $ change according to
\begin{align*}
V &  =V_{0}+\psi_{a}\chi^{a}+\left(  \psi_{a}\xi B_{bc}^{a}+\frac{1}%
{2}\varepsilon_{abc}B^{a}\right)  \chi^{b}\chi^{c}\\
&  +\left(  \varepsilon_{afb}B^{a}\xi B_{cd}^{f}+\frac{1}{3!}\kappa
\varepsilon_{bcd}+\psi_{a}C^{a}\varepsilon_{bcd}\right)  \chi^{b}\chi^{c}%
\chi^{d}%
\end{align*}
The final term vanishes if we choose
\[
\xi B_{bc}^{a}=\frac{\kappa+6\psi_{a}C^{a}}{4B^{2}}\left(  \delta_{b}^{a}%
B_{c}-\delta_{c}^{a}B_{b}\right)
\]
while no choice of $B_{bc}^{a}$ can make the second term vanish because
$\psi_{a}\xi B_{bc}^{a}$ is nilpotent while $\frac{1}{2}\varepsilon_{abc}%
B^{a}$ is not. Renaming the coefficient functions, $V$ takes the form
\[
V\left(  \theta^{b}\right)  =V_{0}+\psi^{a}\theta^{a}+\frac{1}{2}%
\varepsilon_{abc}B^{a}\theta^{b}\theta^{c}.
\]

Now, without loss of generality, the action takes the form%
\[
S=\int dt\left(  \frac{1}{2}m\dot{x}^{i}\dot{x}^{i}+\frac{i}{2}\theta^{a}%
\dot{\theta}^{a}-V_{0}-\psi_{a}\theta^{a}-\frac{1}{2}\varepsilon_{abc}%
B^{a}\theta^{b}\theta^{c}\right)  .
\]
Varying, we get two sets of equations of motion:
\begin{align*}
m\ddot{x}^{i} &  =-\frac{\partial V}{\partial x^{i}}=-\frac{\partial V_{0}%
}{\partial x^{i}}+\frac{\partial\psi^{a}}{\partial x^{i}}\theta^{a}+\frac
{1}{2}\varepsilon_{abc}\frac{\partial B^{a}}{\partial x^{i}}\theta^{b}%
\theta^{c}\\
\dot{\theta}^{a} &  =i\psi^{a}+i\varepsilon_{\ bc}^{a}B^{b}\theta^{c}.
\end{align*}
Clearly this generalizes Newton's second law. The coefficients in the first
equation depend only on $x^{i},$ so terms with different powers of $\theta
^{a}$ must vanish separately. Therefore, $B^{a}$ and $\psi^{a}$ are constant
and we can integrate the $\theta^{a}$ equation immediately. Since $\left[
J_{b}\right]  _{\ c}^{a}=\varepsilon_{cb}^{\ \ a}$ satisfies
\[
\left[  J_{a},J_{b}\right]  _{\ d}^{c}=\varepsilon_{\ ba}^{e}\left[
J_{e}\right]  _{\ d}^{c}%
\]
we see that $B^{b}\varepsilon_{\ bc}^{a}$ is an element of the Lie algebra of
$SO(3)$. Exponentiating to get an element of the rotation group, the solution
for $\theta^{a}$ is
\[
\theta^{a}=i\psi^{a}t+e^{iB^{b}t\varepsilon_{\ bc}^{a}}\theta_{0}^{c}%
\]
The solution for $x^{i}$ depends on the force, $-\partial V/\partial x^{i},$
in the usual way.

It is tempting to interpret the $\theta^{a}$ variables as spin degrees of
freedom and $B^{a}$ as the magnetic field. Then the solution shows that the
spin precesses in the magnetic field. However, notice that $B^{b}%
\varepsilon_{\ bc}^{a}$ is in $SO(3),$ not the spin group $SU(2).$ The
coordinates $\theta^{a}$ therefore provide an example of fermionic, spin-$1$ objects.

One of the goals of early explorations of pseudomechanics was to ask what
classical equations lead to the Pauli and Dirac equations when quantized.
Casalbuoni (\cite{Casalbuoni76a},\cite{Casalbuoni76b}), see also \cite{Freund}
showed how to introduce classical, anticommuting spinors using an
$\hbar\rightarrow0$ limit of a quantum system. Conversely, the action
\[
S=\int dt\left[  \frac{1}{2}m\mathbf{\dot{x}}^{2}+\frac{i}{2}\theta^{a}%
\dot{\theta}^{a}-V_{0}\left(  \mathbf{x}\right)  -\left(  \mathbf{L}%
\cdot\mathbf{S}\right)  V_{LS}-\kappa\frac{1}{2}\left(  \mathbf{S}%
\cdot\mathbf{B}\right)  \right]
\]
where $\mathbf{L}$ is the orbital angular momentum, $\mathbf{S}=-\frac{i}%
{2}\varepsilon_{\ bc}^{a}\theta^{b}\theta^{c},$ and $V_{LS}$ is a spin-orbit
potential, leads to the Pauli-Schr\"{o}dinger equation when quantized.
Similarly, Berezin and Marinov \cite{BerezinMarinov77}, Brink, Deser, Zumino,
and di Vecchia and Howe \cite{BrinkDZdVH},introduced four anticommuting
variables, $\theta^{\alpha}$ to write the pre-Dirac action,
\[
S_{Dirac}=\int d\lambda\left(  -m\sqrt{-v^{\alpha}v_{\alpha}}+\frac{i}%
{2}\left[  \theta_{\beta}\frac{d\theta^{\beta}}{d\lambda}+u_{\alpha}%
\theta^{\alpha}u_{\beta}\frac{d\theta^{\beta}}{d\lambda}-\alpha\left(
u_{\alpha}\theta^{\alpha}+\theta_{5}\right)  \right]  \right)
\]
where
\[
v^{\alpha}=\frac{dx_{\alpha}}{d\lambda},\text{ \ }u^{\alpha}=\frac{v^{\alpha}%
}{\sqrt{-v^{2}}}%
\]
and $\alpha$ is a Lagrange multiplier. The action, $S_{Dirac},$ is both
reparameterization invariant and Lorentz invariant. Its variation leads to the
usual relativistic mass-energy-momentum relation together with a constraint.
When the system is quantized, imposing the constraint on the physical states
gives the Dirac equation.

Evidently, the quantization of these actions is also taken to include the
entension to the relevant covering group.

\subsection{Spin-statistics theorem}

Despite the evident classical independence of spin and statistics, there
exists a limited spin-statistics theorem due to Morgan \cite{Morgan}. The
theorem is proved from Poincar\'{e} invariance, using extensive transcription
of quantum methods into the language of Poisson brackets -- an interesting
accomplishment in itself. A brief statement of the theorem is the following:

\textbf{Theorem}:\textbf{ }Let $L$ be a pseudoclassical,
Poincar\'{e}-invariant Lagrangian, built quadratically from the dynamical
variables. If $L$ is invariant under the combined action of charge conjugation
(C) and time reversal (T) then integer spin variables are even Grassmann
quantities while odd-half-integer spin variables are odd Grassmann quantities.

\textbf{Proof} relies on extending the quantum notions of charge conjugation
and time reversal. As in quantum mechanics, charge conjugation is required to
include complex conjugation. For fermionic variables, Morgan requires reversal
of the order of Grassmann variables under conjugation
\[
\left(  \eta\xi\right)  ^{\ast}=\xi^{\ast}\eta^{\ast}%
\]
This insures the reality property $\left(  \eta\xi^{\ast}\right)  ^{\ast}%
=\eta\xi^{\ast},$ but this is not a necessary condition for complex Grassmann
numbers. For example, the conjugate of the complex $2$-form
\[
\mathbf{d}z\wedge\mathbf{d}z^{\ast}%
\]
is clearly just
\[
\mathbf{d}z^{\ast}\wedge\mathbf{d}z
\]
and is therefore pure imaginary. We must therefore regard the $TC$ symmetry
required by the proof as somewhat arbitrary.

Similarly, for time reversal, \cite{Morgan} requires both
\begin{align*}
t  &  \rightarrow-t\\
\tau &  \rightarrow-\tau
\end{align*}
Whether this is an allowed Poincar\'{e} transformation depends on the precise
definition of the symmetry. If we define Poincar\'{e} transformations as those
preserving the infinitesimal line element, $d\tau,$ then reversing proper time
is not allowed. Of course, we could define Poincar\'{e} transformations as
preserving the quadratic form, $d\tau^{2}=g_{\alpha\beta}dx^{\alpha}dx^{\beta
},$ in which case the transformation is allowed.

Despite its shortcomings, the proof is interesting because it identifies a set
of conditions under which a classical pseudomechanics action obeys the spin
statistics theorem. This is an interesting class of theories and it would be
worth investigating further. Surely there is some set of properties which can
be associated with the classical version of the theorem. Perhaps a fruitful
approach would be to assume the theorem and derive the maximal class of
actions satisfying it.

There are other questions we might ask of spinorial and graded classical
mechanics. A primary question is whether there are any actual physical systems
which are well modeled by either spinors or graded variables. If such systems
exist, are any of them supersymmetric? What symmetries are associated with
spinorial and fermionic variables? Is there a generalization of the Noether
theorem to these variables? What are the resulting conserved quantities? What
is the supersymmetric extension of familiar problems such as the Kepler or
harmonic oscillator?

The statistical behavior of fermionic classical systems is not clear. Quantum
mechanically, of course, Fermi-Dirac statistics follow from the limitation of
discrete states to single occupancy. This, in turn, follows from the action of
an anticommuting raising operator on the vacuum:
\begin{align*}
a^{\dagger}\left|  0\right\rangle  &  =\left|  1\right\rangle \\
a^{\dagger}a^{\dagger}  &  =0
\end{align*}
Since classical states are not discrete, there may be no such limitation. Do
anticommuting classical variables therefore satisfy Bose-Einstein statistics?
If so, how do Fermi-Dirac quantum states become Bose-Einstein in the classical limit?

The introduction of pseudomechanics has led to substantial formal work on
supermanifolds and symplectic supermanifolds. See \cite{Pimentel},
\cite{Carinena} and references therein.

\section{Observations}

Clearly, the field of classical mechanics has no conclusion, and we do not
provide one here. Within each topic we have tried to provide more questions
than answers. However, in the process of collecting these results, we have
observed a few patterns. In closing, we take note of those.

\begin{enumerate}
\item New elements in classical physics work their way into the field from
fundamental research areas, notably quantum field theory and general
relativity. The former has contributed spinors and anticommuting numbers while
the latter lends the tools of differential geometry to the study of symplectic manifolds.

\item Many of the new insights have been seen only in one or two dimensions.
In these cases, it remains an open question whether the properties even exist
in higher dimensions, which higher dimensions, and why in those dimensions.
This applies particularly to the study of inequivalent Lagrangians and Bohmian
quantum mechanics.

\item Comparatively little use is made of the classical physics ArXiv.
Researchers in the area would benefit by using this ready reference tool.

\item Classical mechanics is now strongly influenced by quantum mechanics. In
addition to Bohmian quantum mechanics, which seeks to realize quantum physics
as some sort of classical system, there is phase space quantization, which
accomplishes much the same thing in a different way. An additional approach is
suggested by the gauge theories of Section 6. These programs demonstrate broad
overlap between the classical and quantum worlds.
\end{enumerate}

\appendix

\section{Alternative Lagrangians in higher dimensions}

We show that the Lagrangian
\[
L=v\int^{v}\frac{\alpha\left(  x,\xi,\hat{\theta}\right)  }{\xi^{2}}%
d\xi+f\left(  x,\hat{\theta}\right)
\]
where $\alpha\left(  x,\xi,\hat{\theta}\right)  $ is a time-independent
constant of the motion and $\hat{\theta}_{v}=\dot{x}^{i}/v$ is a unit vector
in the direction of the velocity, solves one of the Euler-Lagrange equations,
\[
\dot{x}^{i}\left(  \frac{d}{dt}\frac{\partial L}{\partial\dot{x}^{i}}%
-\frac{\partial L}{\partial x^{i}}\right)  =0
\]

First notice that
\begin{align*}
\dot{x}_{i}\frac{d}{dt}\left(  \frac{\dot{x}^{i}}{v}\right)   &  =\dot{x}%
_{i}\ddot{x}^{j}\frac{\partial}{\partial\dot{x}^{j}}\left(  \frac{\dot{x}^{i}%
}{v}\right)  =\frac{1}{v}\dot{x}_{i}\ddot{x}^{j}\left(  \delta_{j}^{i}%
-\frac{\dot{x}^{i}\dot{x}_{j}}{v^{2}}\right)  \\
&  =0.
\end{align*}
We therefore have
\begin{align*}
\dot{x}_{i}\left(  \frac{d}{dt}\frac{\partial L}{\partial\dot{x}^{i}}%
-\frac{\partial L}{\partial x^{i}}\right)   &  =\dot{x}_{i}\frac{d}{dt}\left[
\frac{\dot{x}^{i}}{v}\int^{v}\frac{\alpha\left(  x,\xi\right)  }{\xi^{2}}%
d\xi+\frac{\alpha\left(  x,v\right)  }{v}\frac{\dot{x}^{i}}{v}+\frac{\partial
f\left(  \mathbf{x},\vec{\theta}_{v}\right)  }{\partial\dot{x}^{i}}\right]  \\
&  -\dot{x}^{i}v\int^{v}\frac{1}{\xi^{2}}\frac{\partial\alpha\left(
x,\xi\right)  }{\partial x^{i}}d\xi-\dot{x}^{i}\frac{\partial f\left(
\mathbf{x},\vec{\theta}_{v}\right)  }{\partial x^{i}}.
\end{align*}%
\begin{align*}
\dot{x}_{i}\left(  \frac{d}{dt}\frac{\partial L}{\partial\dot{x}^{i}}%
-\frac{\partial L}{\partial x^{i}}\right)    & =\\
\frac{\dot{x}^{i}}{v}\dot{x}_{i}\frac{d}{dt}\int^{v}\frac{\alpha\left(
x,\xi\right)  }{\xi^{2}}d\xi+\frac{\dot{x}^{i}}{v}\dot{x}_{i}\frac{d}{dt}%
\frac{\alpha\left(  x,v\right)  }{v}  & \\
+\dot{x}_{i}\frac{d}{dt}\frac{\partial f}{\partial\dot{x}^{i}}-\dot{x}%
^{i}v\int^{v}\frac{1}{\xi^{2}}\frac{\partial\alpha\left(  x,\xi\right)
}{\partial x^{i}}d\xi-\dot{x}^{i}\frac{\partial f}{\partial x^{i}}  &
\end{align*}%
\begin{align*}
\dot{x}_{i}\left(  \frac{d}{dt}\frac{\partial L}{\partial\dot{x}^{i}}%
-\frac{\partial L}{\partial x^{i}}\right)   &  =\\
v\left(  \ddot{x}_{i}\frac{\dot{x}^{i}}{v}\frac{\alpha\left(  x,v\right)
}{v^{2}}+\dot{x}^{i}\int^{v}\frac{\alpha,_{x^{i}}}{\xi^{2}}d\xi\right)   &  \\
+v\left(  \frac{1}{v}\frac{d\alpha}{dt}-\frac{\alpha\left(  x,v\right)
}{v^{2}}\ddot{x}_{i}\frac{\dot{x}^{i}}{v}\right)   &  \\
-\dot{x}^{i}v\int^{v}\frac{1}{\xi^{2}}\frac{\partial\alpha\left(
x,\xi\right)  }{\partial x^{i}}d\xi+\dot{x}_{i}\frac{d}{dt}\frac{\partial
f}{\partial\dot{x}^{i}}-\dot{x}^{i}\frac{\partial f}{\partial x^{i}} &  \\
=\ddot{x}_{i}\dot{x}^{i}\frac{\alpha}{v^{2}}+v\dot{x}^{i}\int^{v}\frac
{\alpha,_{x^{i}}}{\xi^{2}}d\xi+\frac{d\alpha}{dt}-\frac{\alpha}{v^{2}}\ddot
{x}_{i}\dot{x}^{i} &  \\
-\dot{x}^{i}v\int^{v}\frac{\alpha,_{x^{i}}}{\xi^{2}}d\xi+\dot{x}_{i}\frac
{d}{dt}\frac{\partial f}{\partial\dot{x}^{i}}-\dot{x}^{i}\frac{\partial
f}{\partial x^{i}} &  \\
=\frac{d\alpha}{dt}+\dot{x}_{i}\left(  \frac{d}{dt}\frac{\partial f}%
{\partial\dot{x}^{i}}-\frac{\partial f}{\partial x^{i}}\right)   &  \\
=0 &
\end{align*}
Possibly the function $f$ may be chosen so that the remaining equations of
motion are satisfied.

\section{Arbitrary number of extrema in Kepler orbits}

The problem of global properties of orbits remains open -- power law forces
have been studied \cite{RayShamanna} and found to have limited numbers of
extrema, but non-monotonic force laws allow arbitrarily many extrema. We
provide a simple example here.

Consider the potential
\[
V=\alpha\left(  r-r_{0}\right)  ^{2p}%
\]
The potential $V$ has energy and effective potential,
\[
E=\frac{1}{2}m\dot{r}^{2}+\frac{M^{2}}{2mr^{2}}+\alpha\left(  r-r_{0}\right)
^{2p}%
\]%
\[
V_{eff}=\frac{M^{2}}{2mr^{2}}+\alpha\left(  r-r_{0}\right)  ^{2p}%
\]
This effective potential has an arbitrarily strong minimum near $r_{0}.$ The
exact location of the minimum is given by
\[
0=V_{eff}^{\prime}=2p\alpha\left(  r-r_{0}\right)  ^{2p-1}-\frac{M^{2}}%
{mr^{3}}.
\]
We solve this approximately as follows. Let $r=r_{0}+a.$ Then
\[
0=2p\alpha r_{0}^{3}\left(  1+3\frac{a}{r_{0}}+3\frac{a^{2}}{r_{0}^{2}}%
+\frac{a^{3}}{r_{0}^{3}}\right)  a^{2p-1}-\frac{M^{2}}{m}%
\]
Now suppose $a<<r_{0}$ so that we can neglect the $\frac{a}{r_{0}}$ terms.
Then in order to have solutions we must have
\[
a^{2p-1}=\frac{M^{2}}{2p\alpha mr_{0}^{3}}%
\]
or
\begin{equation}
\left(  \frac{a}{r_{0}}\right)  ^{2p-1}=\frac{M^{2}}{2p\alpha mr_{0}^{2p+2}%
}<<1\label{Inequality}%
\end{equation}
This may be satisfied by choosing $p$ sufficiently large. Thus, $r=r_{0}+a$ is
the approximate position of the extremum. This solution for $r$ is a minimum
since
\[
V_{eff}^{\prime\prime}=\frac{3M^{2}}{mr^{4}}+2p\left(  2p-1\right)
\alpha\left(  r-r_{0}\right)  ^{2p-2}>0
\]
Now, setting $r=r_{0}+a+\varepsilon$, and expanding the effective potential to
second order about the minimum at $r_{0}+a,$%
\begin{align*}
V_{eff} &  =\frac{M^{2}}{2mr_{0}^{2}\left(  1-\frac{a}{r_{0}}+\varepsilon
\right)  ^{2}}+\alpha a^{2p}\left(  1+\frac{\varepsilon}{a}\right)  ^{2p}\\
&  =\frac{M^{2}}{2mr_{0}^{2}}\left[  1+\frac{2a}{r_{0}}-2\varepsilon+\left(
-\frac{a}{r_{0}}+\varepsilon\right)  ^{2}\right]  \\
&  +\alpha a^{2p}\left(  1+\frac{2p\varepsilon}{a}+2p\left(  2p-1\right)
\frac{\varepsilon^{2}}{a^{2}}\right)  \\
&  =\frac{M^{2}}{2mr_{0}^{2}}\left(  1+\frac{2a}{r_{0}}+\frac{a^{2}}{r_{0}%
^{2}}\right)  +\alpha a^{2p}\\
&  -\frac{M^{2}}{mr_{0}^{2}}\left(  1+\frac{a}{r_{0}}\right)  \varepsilon
+2p\alpha a^{2p-1}\varepsilon\\
&  +\frac{M^{2}}{2mr_{0}^{2}}\varepsilon^{2}+2p\left(  2p-1\right)  \alpha
a^{2p-2}\varepsilon^{2}%
\end{align*}
The first term is just an overall constant, while the linear term vanishes
because $r_{0}+a$ is a minimum. The third term is a harmonic oscillator
potential. The approximate equation of motion for the oscillator is
\[
\frac{d^{2}\varepsilon}{dt^{2}}+\frac{1}{m}\left(  \frac{M^{2}}{2mr_{0}^{2}%
}+2p\left(  2p-1\right)  \alpha a^{2p-2}\right)  \varepsilon^{2}=0
\]
By eq.(\ref{Inequality}), the squared frequency becomes
\begin{align*}
\frac{M^{2}}{2m^{2}r_{0}^{2}}+\frac{2p}{m}\left(  2p-1\right)  \alpha a^{2p-2}
&  =\frac{M^{2}}{2m^{2}r_{0}^{2}}+\frac{2p}{m}\left(  2p-1\right)  \alpha
a^{2p-2}\\
&  =\frac{M^{2}}{2m^{2}r_{0}^{2}}+\frac{2p}{ma}\left(  2p-1\right)
\alpha\frac{M^{2}}{2p\alpha mr_{0}^{3}}\\
&  =\frac{M^{2}}{2m^{2}r_{0}^{2}}\left[  1+\frac{2}{ar_{0}}\left(
2p-1\right)  \right]
\end{align*}
The frequency may be made arbitrarily large, at fixed angular momentum $M,$ by
increasing $p.$ This means we may have arbitrarily many extrema per orbit.

\section{The Korteweg-de Vries equation}

While we have so far stayed within particle mechanics, many classical field
theories also have interesting properties. Of particular interest are the
\textquotedblleft integrable systems\textquotedblright\ such as the KdV and
the sine-Gordon equations. These differential equations turn out to have
infinitely many constants of motion.

The KdV equation is a one dimensional, third order field equation that
provides a good example of hidden symmetries. Here we briefly examine some of
its properties. The interesting history of the equation stretches over more
than a century \cite{AbrahamMarsden}. We will begin with a modern form of the
equation,
\[
u_{t}=-6uu_{x}+u_{xxx}%
\]
Consider any function of the form $u=f(z)=f\left(  x-vt\right)  .$
Substituting, we find that $f$ must satisfy
\[
0=\left(  -6f\partial+v\partial+\partial^{3}\right)  f
\]
Let $g=f+c,$ this becomes
\[
0=\left(  -6g\partial+\left(  v+6c\right)  \partial+\partial^{3}\right)  g
\]
so choosing $c=-v/6$ we have simply
\[
0=-6gg_{x}+g_{xxx}%
\]
Integrating twice we find the quadrature,
\[
\int\frac{df}{\sqrt{2af-2f^{3}-vf^{2}+2b}}=x-vt
\]
These solutions for $u$ propagate with unchanging shape $f$ and constant
velocity $v.$ It can be shown that pairs of solitary waves can pass through
one another and emerge unchanged. It has been suggested that the infinite
hierarchy of constants of the motion of this system is related to the
existence of such soliton solutions. Showing how these constants arise will
simultaneously illustrate some techniques of classical field theory. Our
treatment follows that of Abraham and Marsden \cite{AbrahamMarsden}, which is
recommended for further detail.

First, we show that the KdV equation may be described as a Hamiltonian system.
For particle motion expressed in canonical coordinates, we can define the
\textit{Hamiltonian vector field}, $X_{H}$ which is everywhere tangent to the
phase space motion. Conversely, the classical motion of the system is along
the integral curves of this vector field. Restricted to any solution curve,
$X_{H}$ is therefore given by
\begin{align*}
&  \mid X_{H}^{A}\mid_{C=\left(  x\left(  t\right)  ,p\left(  t\right)
\right)  }=\left(  \frac{dx^{i}}{dt},\frac{dp_{i}}{dt}\right) \\
&  =\left(  \frac{\partial H}{\partial p_{j}},-\frac{\partial H}{\partial
x^{i}}\right) \\
&  =\mid\Omega^{AB}\frac{\partial H}{\partial\xi^{B}}\mid_{C=\left(  x\left(
t\right)  ,p\left(  t\right)  \right)  }%
\end{align*}
We therefore can characterize $X_{H}$ everywhere by writing
\[
\frac{\partial H}{\partial\xi^{B}}=\Omega_{BA}X_{H}^{A}%
\]
or more simply using differential forms,
\[
\mathbf{d}H\left(  v\right)  =\mathbf{\omega}\left(  X_{H},v\right)
\]
for any vector field, $v$.

The same relationship holds in classical field theory. For the KdV equation,
we can define a symplectic form as follows:
\[
\omega\left(  u,v\right)  =\frac{1}{2}\int dx\int^{x}dy\left[  u\left(
y\right)  v\left(  x\right)  -u\left(  x\right)  v\left(  y\right)  \right]
\]
Here $u$ and $v$ are arbitrary vector fields. Now suppose the Hamiltonian is
given as an integral over a Hamiltonian density,
\[
H=\int f[u\left(  x\right)  ]dx
\]
Hamilton's equations are then involve functional derivatives. For the
differential of the Hamiltonian,
\[
\mathbf{d}H\left(  v\right)  =\int_{-\infty}^{\infty}dx\frac{\delta f}{\delta
u}\left(  x\right)  v\left(  x\right)
\]
so equating to the symplectic form,
\begin{align*}
\mathbf{d}H\left(  v\right)   &  =\mathbf{\omega}\left(  X_{H},v\right) \\
\int_{-\infty}^{\infty}dx\frac{\delta f}{\delta u}\left(  x\right)  v\left(
x\right)   &  =\frac{1}{2}\int dx\int^{x}dy\left[  X_{H}\left(  y\right)
v\left(  x\right)  -X_{H}\left(  x\right)  v\left(  y\right)  \right]
\end{align*}
We seek an expression for $X_{H}.$ First, write $X_{H}$ in the form
\[
X_{H}=\frac{\partial G}{\partial x}%
\]
Then, integrating by parts and disregarding surface terms, we obtain,
\begin{align*}
\int_{-\infty}^{\infty}dx\frac{\delta f}{\delta u}\left(  x\right)  v\left(
x\right)   &  =\frac{1}{2}\int dx\int^{x}dy\left[  \frac{\partial G}{\partial
y}v\left(  x\right)  -\frac{\partial G}{\partial x}v\left(  y\right)  \right]
\\
&  =\frac{1}{2}\int dxv\left(  x\right)  \int^{x}dy\frac{\partial G}{\partial
y}-\frac{1}{2}\int dx\frac{\partial G}{\partial x}\int^{x}dyv\left(  y\right)
\\
&  =\int dxv\left(  x\right)  G\left(  x\right)
\end{align*}
and since $v$ is arbitrary we have
\[
G\left(  x\right)  =\frac{\delta f}{\delta u}\left(  x\right)
\]
Therefore,
\[
X_{H}=\frac{\partial}{\partial x}\frac{\delta f}{\delta u}.
\]

Now consider the KdV equation,
\[
u_{t}=6uu_{x}-u_{xxx}%
\]
The time evolution of $u$ is given by the tangent vector field $u_{t}.$ We ask
if we can write this vector field as a Hamiltonian vector field $X_{H}$ for
some Hamiltonian $H.$ Equating
\[
u_{t}=X_{H}%
\]
we require
\begin{align*}
X_{H}  &  =\frac{\partial}{\partial x}\frac{\delta f}{\delta u}=6uu_{x}%
-u_{xxx}=\frac{\partial}{\partial x}\left(  3u^{2}-u_{xx}\right) \\
\frac{\delta f}{\delta u}  &  =3u^{2}-u_{xx}%
\end{align*}
and it is easy to see that we can take
\begin{align*}
f  &  =u^{3}+\frac{1}{2}u_{x}^{2}\\
H  &  =\int dx\left(  u^{3}+\frac{1}{2}u_{x}^{2}\right)  .
\end{align*}
We therefore have a Hamiltonian system, and the KdV equation may be studied in
terms of a Hamiltonian flow.

We now show that the KdV equation possesses infinitely many constants of
motion. Define an infinite set of Hamiltonian vector fields and Hamiltonian
densities by acting repeatedly on $X_{1}$ and $f_{1}$ according to
\begin{align*}
X_{n+1} &  =\left(  2au\partial_{x}+au_{x}+b\partial_{x}^{3}\right)
\frac{\delta f_{n}}{\delta u}\\
\frac{\partial}{\partial x}\frac{\delta f_{n+1}}{\delta u} &  =X_{n+1}%
\end{align*}
The first expression always exists, but the second is possible as long as each
new $X_{n+1}$ is a Hamiltonian flow. In order for there to exist a Hamiltonian
such that
\[
\mathbf{d}H\left(  v\right)  =\mathbf{\omega}\left(  X_{H},v\right)
\]
we require the integrability condition
\[
0\equiv\mathbf{d}^{2}H\left(  v\right)  =\mathbf{d\omega}\left(
X_{H},v\right)
\]
Essentially, this condition reduces to the equality of mixed partial
functional derivatives. We omit the inductive proof that shows that the
condition is satisfied for all $X_{n},$ as long as it holds for the initial
set. Letting $f_{1}=u^{2}/2$ it follows that
\begin{align*}
X_{2} &  =\partial_{x}\left(  3u^{2}-u_{xx}\right)  =6uu_{x}-\partial_{x}%
^{3}u\\
\frac{\partial}{\partial x}\frac{\delta f_{2}}{\delta u} &  =\partial
_{x}\left(  3u^{2}-u_{xx}\right)  \\
\frac{\delta f_{2}}{\delta u} &  =3u^{2}-u_{xx}\\
f_{2} &  =u^{3}+\frac{1}{2}u_{x}^{2}%
\end{align*}
so $f_{2}$ is the Hamiltonian density for the KdV equation,
\[
H=\int dx\left(  u^{3}+\frac{1}{2}u_{x}^{2}\right)
\]
Since the inductive hypothesis holds, the entire set of Hamiltonian vector
fields $X_{n}$ and Hamiltonian densities $f_{n}$ exists.

Finally, we are ready for the proof that there exist an infinite number of
constants of the motion of the KdV equation. Consider the higher order
\textquotedblleft Hamiltonians\textquotedblright\ given by integrating the
$f_{n}:$%
\[
H_{n}=\int f_{n}\left(  x\right)  dx
\]
We compute their Poisson brackets with one another by integration by parts,
\begin{align*}
\left\{  H_{n},H_{m}\right\}   &  =\Omega\left(  X_{n},X_{m}\right) \\
&  =\frac{1}{2}\int dx\int^{x}dy\left[  X_{n}\left(  y\right)  X_{m}\left(
x\right)  -X_{n}\left(  x\right)  X_{m}\left(  y\right)  \right] \\
&  =\frac{1}{2}\int dx\int^{x}dy\left[  \partial_{y}\frac{\delta f_{n}}{\delta
u}\left(  y\right)  X_{m}\left(  x\right)  -\partial_{x}\frac{\delta f_{n}%
}{\delta u}\left(  x\right)  X_{m}\left(  y\right)  \right] \\
&  =\int dx\frac{\delta f_{n}}{\delta u}X_{m}\\
&  =\int dx\frac{\delta f_{n}}{\delta u}\left(  2au\partial_{x}+au_{x}%
+b\partial_{x}^{3}\right)  \frac{\delta f_{m-1}}{\delta u}\\
&  =\int dx\left(  2a\frac{\delta f_{n}}{\delta u}u\partial_{x}\frac{\delta
f_{m-1}}{\delta u}+a\frac{\delta f_{n}}{\delta u}u_{x}\frac{\delta f_{m-1}%
}{\delta u}+b\frac{\delta f_{n}}{\delta u}\partial_{x}^{3}\frac{\delta
f_{m-1}}{\delta u}\right) \\
&  =\int dx\left[  -2a\partial_{x}\left(  u\frac{\delta f_{n}}{\delta
u}\right)  \frac{\delta f_{m-1}}{\delta u}\right] \\
&  +\int dx\left(  au_{x}\frac{\delta f_{n}}{\delta u}\frac{\delta f_{m-1}%
}{\delta u}-b\partial_{x}^{3}\frac{\delta f_{n}}{\delta u}\frac{\delta
f_{m-1}}{\delta u}\right) \\
&  =-\int dx\frac{\delta f_{m-1}}{\delta u}\left(  au_{x}+2au\partial
_{x}+b\partial_{x}^{3}\right)  \frac{\delta f_{n}}{\delta u}\\
&  =-\int dx\frac{\delta f_{m-1}}{\delta u}X_{n+1}\\
&  =-\left\{  H_{m-1},H_{n+1}\right\} \\
&  =\left\{  H_{n+1},H_{m-1}\right\}
\end{align*}
Now iterate this relationship. First suppose $n$ and $m$ are either both even
or both odd. Then without loss of generality we take $m-n=2k>0.$ Setting
$m=n+2k$ and iterating $k$ times we have:
\begin{align*}
\left\{  H_{n},H_{m}\right\}   &  =\left\{  H_{n},H_{n+2k}\right\} \\
&  =\left\{  H_{n+k},H_{n+2k-k}\right\} \\
&  =\left\{  H_{n+k},H_{n+k}\right\} \\
&  =0
\end{align*}
where the last step follows by antisymmetry of the bracket. Now let
$m=n+2k+1.$ Again iterating $k-1$ times, and then one more time, give
\begin{align*}
\left\{  H_{n},H_{m}\right\}   &  =\left\{  H_{n},H_{n+2k+1}\right\} \\
&  =\left\{  H_{n+k-1},H_{n+k}\right\} \\
&  =\left\{  H_{n+k},H_{n+k-1}\right\}
\end{align*}
But the last two lines are negatives of one another, and therefore vanish.
Therefore, all of the $H_{n}$ have vanishing Poisson brackets with one
another. In particular, since $H_{2}$ is the original Hamiltonian, $\left\{
H_{2},H_{m}\right\}  =0$ for all $m,$ and the evolution generated by $H_{2}$
leaves all $H_{m}$ constant. Since the evolution by $H_{2}$ generates
solutions to the KdV equation, all $H\,_{n}$ are constants of integration of
the KdV system.

The KdV equation has interesting quantum properties as well. It can be shown
that the Schr\"{o}dinger equation with time-dependent potential $u\left(
t\right)  $ has solutions with a fixed energy spectrum -- the time-dependence
of the potential does not change the energies of the soltutions. The proof
hinges on the Lax theorem, which states that the KdV equation is equivalent to
the equation
\[
u_{t}=\left[  L,A\right]
\]
where
\[
A=4\partial_{x}^{3}+6u\partial_{x}+3u_{x}%
\]
and
\[
L=\partial_{x}^{2}+u
\]
This latter operator $L$ is just the Schr\"{o}dinger operator with potential
$u.$ The proof of the theorem follows by direct calculation:
\begin{align*}
u_{t}f  &  =\left[  4\partial_{x}^{3}+6u\partial_{x}+3u_{x},\partial_{x}%
^{2}+u\right]  f\\
&  =\left(  u_{xxx}+6uu_{x}\right)  f
\end{align*}
This is satisfied if $-u$ satisfies the KdV equation. The full proof of the
resulting isospectral theorem may be found in \cite{AbrahamMarsden}. In light
of this relationship between a remarkable classical system and its equally
striking quantum properties, one wonders whether the relationship between the
classical and quantum mechanics may be much like the relationship between the
real line and the complex plane. Just as real functions often display their
full character only when analytically extended to the complex plane, many
classical systems may show their true natures when quantized. The KdV equation
provides an excellent example of this -- solutions to the KdV equation, when
used as quantum potentials, are isospectral despite time-dependent potentials,
and there may be a profound connection between the KdV and Schr\"{o}dinger systems.

The existence of equations such as the KdV equation, which have infinitely
many independent conserved quantities is remarkable in several regards. For
example, Goldstein, Poole and Safko observe that the KdV equation provides a
counterexample to the converse of the Noether theorem \cite{GoldsteinPS}.
Thus, while symmetries of an action lead to conserved quantities, the KdV and
other equations have infinitely many conserved quantities without
corresponding symmetries.

There is little systematic theory of these so-called ``integrable systems.''
In fact, we lack even a clear definition of this concept of integrability.
Still, there has considerable recent progress (see, for example,
\cite{TalukdarGS} and references therein).

A related question is whether such systems exist in higher dimensions. As with
many modern results in classical mechanics, examples are limited to one or two
dimensions, and it is unclear whether we are seeing properties of geometries
or only of the real and complex number systems.

\section{Chaos}

While the study of nonlinear and chaotic systems is beyond the scope of this
review, one common example provides an interesting case of a higher order
differential equation for a classical system. The onset of chaos may be
visualized by studying the fixed points of the logistic equation,
\[
x_{k+1}=ax_{k}\left(  1+bx_{k}\right)
\]
for varying values of the parameters $a$ and $b$ (see the review articles by
May \cite{May76}, May and Oster \cite{MayOster}, as well as May \cite{May74} )
As the values of these parameters change, the number of fixed points passes
through bifurcation points, leading to more and more frequent doubling of the
their number. At a finite value of the parameters, the number of fixed points
diverges and the behavior of the system is said to become chaotic.

This equation may be converted into a nonlocal equation of continuous motion
for a one dimensional system. The fixed points of the discrete equation then
become periodic solutions for the continuous system. Replace the discrete
sequence $x_{k}$ with a function $x\left(  t\right)  ,$ which must satisfy
\[
x\left(  t+1\right)  =ax\left(  t\right)  [1+bx\left(  t\right)  ]
\]
The left hand side may be expanded in a Taylor series as
\[
x\left(  t+1\right)  =\sum_{n=0}^{\infty}\frac{1}{n!}\mid\frac{d^{n}x}{dt^{n}%
}\mid_{t}\cdot\left(  1\right)  ^{n}%
\]
so at any time $t,$ the function $x$ must satisfy
\[
\sum_{n=0}^{\infty}\frac{1}{n!}\frac{d^{n}x}{dt^{n}}-ax-abx^{2}=0
\]
This certainly qualifies as a higher order differential equation!

We can find the fixed points from the continuous representation as well as
from the discrete one. At $k^{th}$-order fixed points of the discrete system,
we require periodicity of the form $x\left(  t+k\right)  =x\left(  t\right)
.$ To examine the consequences of this condition, we employ a common technique
for periodic systems \cite{LandauLifshitz}.

Suppose $x_{i}(t)$ are independent solutions to $x\left(  t+k\right)
=x\left(  t\right)  .$ Then a general solution may be written as a
superposition of these, so $x_{i}\left(  t+k\right)  $ must be some
superposition:
\[
x_{i}\left(  t+k\right)  =\sum_{j}a_{ij}x_{j}\left(  t\right)
\]
Periodic solutions then satisfy
\[
x_{i}\left(  t\right)  =\sum_{j}a_{ij}x_{j}\left(  t\right)
\]
Now diagonalize $a_{ij}.$ If $y_{i}$ are the new basis functions and
$\lambda_{i}$ the eigenvalues, then
\[
y_{i}\left(  t+k\right)  =\lambda_{i}y_{i}\left(  t\right)
\]
implies
\[
y_{i}\left(  t\right)  =\lambda_{i}^{t/k}\pi_{i}(t)
\]
where $\pi_{i}$ is any periodic function with period $k.$ Now, the $y_{i}$
satisfy the equations
\begin{align*}
\sum_{n=0}^{\infty}\frac{1}{n!}\frac{d^{n}y_{i}}{dt^{n}}-ay_{i}-aby_{i}^{2}
&  =0\\
\sum_{n=0}^{\infty}\frac{k^{n}}{n!}\frac{d^{n}y_{i}}{dt^{n}}  &  =\lambda
_{i}y_{i}\left(  t\right)
\end{align*}
Consider the long-term behavior of $y_{i}.$ Suppose $\lambda_{i}>1$ so that
$y_{i}(t)$ diverges at late times. From the first equation $y_{i}$ must
satisfy
\[
-aby_{i}^{2}=0
\]
as $t\rightarrow\infty,$ so the limiting value of $y_{i}$ is zero, which is
inconsistent. Therefore, we require $\lambda_{i}<1$ so that $y_{i}$ is also
converging to zero at late times, and must approximately satisfy both
\begin{align*}
\sum_{n=0}^{\infty}\frac{1}{n!}\frac{d^{n}y_{i}}{dt^{n}}  &  =ay_{i}\left(
t\right) \\
\sum_{n=0}^{\infty}\frac{k^{n}}{n!}\frac{d^{n}y_{i}}{dt^{n}}  &  =\lambda
_{i}y_{i}\left(  t\right)
\end{align*}
Since these are linear, we may write
\[
y_{i}=e^{\alpha_{i}t}%
\]
Then
\begin{align*}
\sum_{n=0}^{\infty}\frac{\left(  \alpha_{i}\right)  ^{n}}{n!}  &  =a\\
\sum_{n=0}^{\infty}\frac{\left(  \alpha_{i}k\right)  ^{n}}{n!}  &
=\lambda_{i}%
\end{align*}
and we need both
\begin{align*}
e^{\alpha_{i}}  &  =a\\
e^{k\alpha_{i}}  &  =\lambda_{i}=a^{k}%
\end{align*}
thereby determining the (asymptotic) eigenvalues.

\end{document}